\documentclass{article}
\usepackage{bm}
\usepackage{amsmath}
\usepackage{amssymb}
\usepackage[onehalfspacing]{setspace}
\usepackage[utf8]{inputenc}
\usepackage[T1]{fontenc}
\usepackage[hidelinks]{hyperref}
\usepackage{graphicx}
\usepackage{tikz}
\usepackage{algorithm}
\usepackage{algorithmic}
\usepackage{csquotes}
\usepackage{geometry}
\usepackage[numbers]{natbib}
\usepackage{microtype}
\usepackage{cleveref}
\usepackage{newtxtext,newtxmath}
\usepackage{siunitx}
\usepackage{booktabs}
\usepackage{pgfplots}
\pgfplotsset{compat=1.18}
\usetikzlibrary{patterns}
\usepackage{xcolor}
\usepackage{subcaption}
\usepackage{authblk}

\newcommand{\rmd}{\mathrm{d}}
\newcommand{\rmt}{\mathrm{T}}
\newcommand{\bd}{\bm}

\title{Multi-Revolution Low-Thrust Trajectory Optimization With Very Sparse Mesh Pseudospectral Method}
\author{Yilin Zou\thanks{Ph.D. candidate, School of Aerospace Engineering, zouyl22@mails.tsinghua.edu.cn.} }
\author{Fanghua Jiang\thanks{Associate Professor, School of Aerospace Engineering, jiangfh@tsinghua.edu.cn (Corresponding Author).}}
\affil{Tsinghua University, Beijing, China}
\date{\today}

\begin{document}
\maketitle

\begin{abstract}
    Multi-revolution low-thrust trajectory optimization problems are important and challenging in space mission design. In this paper, an efficient, accurate, and widely applicable pseudospectral method is proposed to solve multi-revolution low-thrust trajectory optimization problems with various objective functions and perturbations. The method is based on the Sundman transformation and pseudospectral method, together with a sparse mesh that is monotonic, near-uniformly spaced, and uniformly scattered on the unit circle. Two methods are proposed to construct the mesh: a deterministic method based on rotation mapping; a stochastic method utilizing autocorrelated random sequences. Core mechanisms ensuring the correctness of the method are analyzed, including the dual roles of mesh points as both integration points in the temporal domain and sampling points in the angular domain, the slow dynamics of the system excluding the fast angle variable, and the nearly commutative vector fields generated by applying different control inputs. The method is demonstrated through a multi-revolution low-thrust orbital rendezvous problem. Results show that the proposed method achieves high accuracy with only a few seconds of computational time for challenging problems. 
\end{abstract}

\section{Introduction}
Low-thrust propulsion systems have emerged as transformative technologies in the field of space exploration and satellite operations. These systems utilize electric or ionized propellants to generate thrust and typically achieve a higher specific impulse than traditional chemical propulsion systems, at the cost of lower thrust levels~\cite{moranteSurveyLowThrustTrajectory2021}. The high specific impulse of low-thrust propulsion systems enables more efficient fuel consumption, making them particularly well suited for long-duration missions and deep-space exploration.

Optimal control theory is commonly employed to design low-thrust trajectories, with the goal of minimizing a cost function such as fuel consumption, time of flight, or other mission-specific objectives. The trajectory must satisfy the spacecraft's dynamics, path constraints, and boundary conditions. Optimal control problems are typically solved using numerical methods, which can be categorized into direct and indirect methods~\cite{bettsSurveyNumericalMethods1998}. Indirect methods transform the optimal control problem into a two-point boundary value problem (TPBVP) using calculus of variations and Pontryagin's maximum principle (PMP). The TPBVP is then solved using numerical methods, typically based on the shooting method with homotopy approaches, to cope with discontinuities in the control inputs~\cite{bertrandNewSmoothingTechniques2002,haberkornLowThrustMinimumFuel2004,jiangPracticalTechniquesLowThrust2012,wuWarmStartMultihomotopicOptimization2020,ottesenDirecttoindirectMappingOptimal2024}.

Direct methods discretize the continuous problem into a finite-dimensional nonlinear programming problem and then solve it using optimization solvers designed for large, sparse problems~\cite{topputoSurveyDirectTranscription2014}. Pseudospectral methods are particularly popular among direct methods due to their theoretical efficiency in function interpolation and numerical integration~\cite{rossReviewPseudospectralOptimal2012}. In these methods, state and control variables are discretized using interpolating polynomials based on carefully chosen interpolation nodes within each subinterval, with the entire time domain divided into multiple subintervals. The dynamics and path constraints for the state and control variables are discretized as constraints imposed at the collocation points, which are set to coincide with the interpolation nodes. The objective function is approximated using Gaussian quadrature, thereby transforming the continuous problem into a finite-dimensional nonlinear programming (NLP) problem. The resulting NLP problem typically involves a large number of variables and constraints, and can be solved using optimization solvers such as IPOPT~\cite{wachterImplementationInteriorpointFilter2006}, SNOPT~\cite{gillSNOPTSQPAlgorithm2002}, and Knitro~\cite{byrdKnitroIntegratedPackage2006a}. In pseudospectral methods, interpolation nodes are generally chosen as the roots of orthogonal polynomials to maximize the accuracy of polynomial approximations and numerical integration. Common choices include Legendre-Gauss (LG)~\cite{LG1,LG2}, Legendre-Gauss-Radau (LGR)~\cite{LGR1,LGR2}, and Legendre-Gauss-Lobatto (LGL) points~\cite{LGL1,LGL2}. All these nodes are derived from Legendre polynomials; however, they differ in whether endpoints are included and in the integration order~\cite{gargUnifiedFrameworkNumerical2010}. Typical implementations of the pseudospectral methods include GPOPS-II~\cite{pattersonGPOPSIIMATLABSoftware2014a} and DIDO~\cite{rossEnhancementsDIDOOptimal2020}.

However, both indirect and direct methods face challenges when applied to trajectory optimization problems with low-thrust and multi-revolution characteristics. For indirect methods, long-duration missions with low-thrust propulsion systems cause the final values of the variables to be highly sensitive to the initial guess, creating substantial difficulties for the shooting method's convergence~\cite{haberkornLowThrustMinimumFuel2004}. For direct methods, capturing spacecraft dynamics typically requires a large number of collocation points, leading to numerous variables and constraints in the resulting nonlinear programming problem. As an example, Betts solved a \num{578}-revolution trajectory optimization problem with \num{416123} variables and \num{249674} constraints~\cite{bettsVeryLowthrustTrajectory2000}, which required thousands of seconds of computational time. This results in high computational costs, especially for multi-revolution problems where trajectories may involve hundreds or thousands of revolutions. Additionally, discontinuities in the control inputs, inherent in bang-bang control for fuel-optimal problems, cause numerical difficulties in function approximation and optimization~\cite{hagerConvergenceRateGauss2016}.

Many efforts have been made to improve the efficiency of trajectory optimization for low-thrust propulsion systems. Among these efforts, the averaging method is particularly important and effective. The averaging method originates from the key observation that the dynamical system of a low-thrust trajectory optimization problem can be characterized by a set of slow variables and a fast angle variable. The slow variables characterize the spacecraft's orbit and mass, while the fast variable is the angle describing the specific position of the spacecraft on its orbit. In a low-thrust trajectory optimization problem, the slow variables change at rates orders of magnitude slower than the fast variable. The averaging method averages the system over one revolution of the angle variable, thus creating an averaged system that contains only the slow variables. The averaged system still captures the core characteristics of the original system but can be solved analytically or numerically with a much larger time step. Edelbaum et al.~\cite{edelbaumPropulsionRequirementsControllable1961} and related works~\cite{kechichianReformulationEdelbaumsLowThrust1997} focused on minimum-time low-thrust transfers between circular orbits. The averaging method was used to derive an analytical solution for the secular variations. Bonnard et al.~\cite{bonnardRiemannianMetricAveraged2007,bonnardGeometricOrbitalTransfer2008,bonnardGeodesicFlowAveraged2009} conducted a series of studies on the geometric properties of the averaged system for minimum-energy problems, remarkably solving the BVP of the averaged system analytically for the coplanar case. Another class of methods, as in~\cite{gaoNearOptimalVeryLowThrust2007,huangMixedOptimizationApproach2024}, divides the trajectory into several stages with prescribed control laws, and then uses the averaging method to solve the system effectively for each stage. Wu et al.~\cite{wuLowThrustTrajectoryOptimization2024} extended the idea of the averaging method to minimum-time and minimum-fuel problems, integrating it with other techniques such as neural networks.

One difficulty of the averaging method is that for many combinations of objective functions, dynamics, and path constraints, the averaged system cannot be analytically solved in closed form, limiting its application. Absorbing the core ideas of the averaging method, such as the separation of fast and slow variables, this paper applies these concepts within a direct pseudospectral method. The approach combines the advantages of both the averaging and pseudospectral methods. It integrates the system with a sparse mesh and achieves high accuracy with a small number of collocation points. This is accomplished by using a sparse mesh while effectively sampling the fast angle variable through a specific choice of mesh points. As a result, the computational cost is significantly reduced, enabling the method to solve trajectory optimization problems with a large number of revolutions. It also possesses the advantages of avoiding complex function derivation and can be easily implemented and applied to various low-thrust trajectory optimization problems.

The remainder of this paper is organized as follows. In Section~\ref{sec:problem}, the multi-revolution low-thrust trajectory optimization problem is precisely formulated. Section~\ref{sec:method} presents the developed methods, including the Sundman transformation, pseudospectral method, and the two sparse mesh construction methods. Theoretical justification of the proposed method is provided in Section~\ref{sec:rationale}, explaining the mechanisms behind its effectiveness. Section~\ref{sec:example} demonstrates the method's performance through a challenging numerical example, highlighting the effects of sparse mesh construction. Conclusions are presented in Section~\ref{sec:conclusion}.

\section{Multi-Revolution Low-Thrust Trajectory Optimization Problem}\label{sec:problem}
The low-thrust orbit optimization problem involves finding the optimal control inputs that minimize the cost function while satisfying the spacecraft's dynamics, path constraints, and boundary conditions.
We adopt the modified equinoctial elements (MEE) $[p, f, g, h, k, L]^\rmt$ as the state variables, where $p$ is the semi-latus rectum, $f$ and $g$ are the components of the eccentricity vector, $h$ and $k$ are the components of the inclination vector, and $L$ is the true longitude. 
Their relationship with the classical orbital elements and the Descartes coordinates can be found in literature~\cite{walkerSetModifiedEquinoctial1985}. 
In addition, the spacecraft's mass $m$ is included as a state variable, so that the system comprises a total of $7$ state variables $\bd{x} = [p, f, g, h, k, L, m]^\rmt$. The control inputs are the components of the thrust vector $\bd{u} = [u_r, u_t, u_n]^\rmt$, where $u_r$ is the radial component, $u_t$ is the transverse component, and $u_n$ is the normal component. The dynamics of the spacecraft is described by the following differential equations~\cite{walkerSetModifiedEquinoctial1985}
\begin{align}
    \dot{p} &= \frac{2 p a_t}{w} \sqrt{\frac {p}{\mu}}\label{eq:dotp}\\
    \dot{f} &= \sqrt{\frac{p}{\mu}} \left\{  a_r \sin L +\frac{\left[(w+1)\cos L+f\right]a_t}{w} - \frac{g(h\sin L-k\cos L)a_n}{w}\right\}\\
    \dot{g} &= \sqrt{\frac{p}{\mu}} \left\{  -a_r \cos L +\frac{\left[(w+1)\sin L+g\right]a_t}{w} + \frac{f(h\sin L-k\cos L)a_n}{w}\right\}\\
    \dot{h} &= \sqrt{\frac{p}{\mu}} \frac{s^2a_n}{2w}\cos L\\
    \dot{k} &= \sqrt{\frac{p}{\mu}} \frac{s^2a_n}{2w}\sin L\\
    \dot{L} &= \sqrt{\mu p} \left(\frac wp\right)^2+ \sqrt{\frac{p}{\mu}} \frac{(h\sin L-k\cos L)a_n}{w}\label{eq:dotL}\\
    \dot{m} &= -\frac {\|\bd{u}\|}{I_\text{sp} g_0}
\end{align}
where $a_r = u_r / m$, $a_t = u_t / m$, $a_n = u_n / m$, $w = 1 + f \cos L + g \sin L$, $s^2 = 1+h^2+k^2$, $\mu$ is the gravitational parameter of the central body, $I_\text{sp}$ is the specific impulse of the thruster, and $g_0$ is the standard gravity acceleration equal to \qty{9.80665}{\metre/\second\squared}. 

For a low-thrust orbital transfer problem, the control inputs are small, and we suppose the magnitute of the control inputs is constrained by
\begin{equation}
    \|\bd{u}\| \leq u_\text{max} \ll \frac{\mu m}{p^2}
\end{equation}
That is, the thrust is much smaller than the gravitational force exerted by the central body.
The cost function is the integral
\begin{equation}\label{eq:objective}
    J = \int_{t_0}^{t_f} F\left(\|\bd{u}\|\right) \rmd t
\end{equation}
The integrand $F$ is required to be a continuous function of the magnitude of the control inputs $\|\bd{u}\|$. Common choices include $F = 1$ for minimum-time problems, $F = \|\bd{u}\|$ for minimum-fuel problems, and $F = \|\bd{u}\|^2$ for minimum-energy problems.

The initial and final values of the state variables, along with the start and end times, can be specified as boundary conditions. These parameters may be either fixed or free in the optimization problem.

In addition to the gravitational force exerted by the central body, additional orbital perturbations, such as the non-spherical Earth gravity or atmospheric drag, may be incorporated by introducing extra terms into the acceleration components $a_r$, $a_t$, and $a_n$. The proposed method remains valid when these perturbations are sufficiently small and exhibit periodic behavior with respect to the true longitude $L$. As such, the primary theoretical arguments are unaffected by their inclusion, and the pertinent equations are not shown here for brevity. Nonetheless, the $J_2$ perturbation is explicitly incorporated in the numerical example to highlight the method's capability for handling such perturbations.

\section{Pseudospectral Method with Sparse Mesh}\label{sec:method}
\subsection{Sundman Transformation and Pseudospectral Method}\label{sec:sundman}
Since the control thrust is small, the true longitude $L$ increases monotonically with the time (see Equation~\eqref{eq:dotL}). This justifies replacing the time $t$ with the true longitude $L$ as the independent variable, a transformation known in the literature as the Sundman transformation. The spacecraft dynamics in the transformed problem are described by the following differential equations
\begin{equation}
    \frac{\rmd p}{\rmd L} = \frac{\dot{p}}{\dot{L}}, \quad
    \frac{\rmd f}{\rmd L} = \frac{\dot{f}}{\dot{L}}, \quad
    \frac{\rmd g}{\rmd L} = \frac{\dot{g}}{\dot{L}}, \quad
    \frac{\rmd h}{\rmd L} = \frac{\dot{h}}{\dot{L}}, \quad
    \frac{\rmd k}{\rmd L} = \frac{\dot{k}}{\dot{L}}, \quad
    \frac{\rmd t}{\rmd L} = \frac{1}{\dot{L}}
\end{equation}
The objective function becomes the integral of $F$ over the true longitude $L$
\begin{equation}
    J = \int_{L_0}^{L_f} \frac{F}{\dot{L}} \rmd L
\end{equation}
where $L_0$ and $L_f$ denote the initial and final true longitudes, respectively. The path constraints and boundary conditions remain unchanged and are applied to the corresponding variables.

The pseudospectral method is applied to the transformed problem by discretizing the continuous optimal control problem into a nonlinear programming problem. Specifically, we evaluated both the Legendre-Gauss-Radau (LGR) pseudospectral method and the Legendre-Gauss-Lobatto (LGL) pseudospectral method in integral form~\cite{gargUnifiedFrameworkNumerical2010}. Our numerical experiments demonstrate that a discretization with a uniform dense mesh (exceeding 10 subintervals per orbital revolution) combined with low-order collocation (2 to 4 points) produces solutions of high accuracy. We observe that the resulting nonlinear programming problem exhibits robust convergence and can be solved within computationally tractable timeframes, typically ranging from a few seconds to a few minutes, depending on the problem size and complexity. Concrete examples implementing the LGL method will be presented in Section \ref{sec:dense_mesh_solution}.

\subsection{Dual Interpretations of Mesh Points}\label{sec:dual_interpretation}
Throughout this article, the term “unit circle” refers to the quotient space $\mathbb{R}/2\pi\mathbb{Z}$, where numbers are considered equivalent if they differ by an integer multiple of $2\pi$. The mesh points, which represent specific values of the true longitude $L$, have two interpretations:
\begin{enumerate}
     \item As elements of the real line $\mathbb{R}$. In this interpretation, $L$ and $L+2\pi k$ are considered distinct when $k\neq 0$. After applying the Sundman transformation, the true longitude $L$ serves as the independent variable. The dynamical system is numerically integrated over $L$ within the interval $[L_0, L_f]$, treating $L$ as a real number. This is analogous to integrating any dynamical system over a real-valued independent variable.
        \item As their canonical projections onto the unit circle $\mathbb{R}/2\pi\mathbb{Z}$, which maps each value to its equivalence class modulo $2\pi$. With this interpretation, $L$ and $L+2\pi k$ are considered identical for any integer $k$. This perspective is important for averaging the system over revolutions of the true longitude $L$. Since the dynamical equations (\ref{eq:dotp}--\ref{eq:dotL}) are periodic with respect to the fast angle variable $L$, the projection of the mesh points onto the unit circle represents sampling the system dynamics at specific angles. This concept is fundamental to the methodology developed in this work and will be discussed in detail in the following sections.
\end{enumerate}
Both interpretations are fundamental to the methodology developed in this work. In subsequent sections, we may omit explicit projection notation for brevity. The phrase “mesh point $L_i$ on the unit circle” implicitly refers to the equivalence class $\{L_i + 2\pi k : k \in \mathbb{Z}\}$, which can be conceptually visualized as points on a circle of unit radius with angular coordinates given by $L_i$.

\subsection{Conditions on Mesh Points}\label{sec:mesh_points}
Suppose the pseudospectral scheme contains $N$ subintervals, with the mesh points denoted as $L_0, L_1, \dots, L_N=L_f$, where $L_0$ and $L_f$ are the initial and final true longitudes, respectively. The following three conditions are found to be essential for the convergence and accuracy of the optimization algorithm:
\begin{enumerate}
    \item Monotonicity: $L_i < L_{i+1}$ for $i = 0, 1, \dots, N-1$ to maintain the stability of numerical integration. In contrast, non-monotonic mesh points would correspond to integrating the system backward in certain intervals, leading to negative time steps and integration weights in pseudospectral methods, thereby producing unfavorable numerical properties.
    \item Near-uniform spacing: The difference between adjacent mesh points should not deviate significantly from $h = \tfrac{L_f - L_0}{N}$, with the optimal configuration being $L_{i+1} - L_i = h$ for all $i = 0, 1, \dots, N-1$. This condition employs the first interpretation of the mesh points as elements of the real line $\mathbb{R}$ as described in Section \ref{sec:dual_interpretation}.
    \item Even sampling: Employing the second interpretation of the mesh points as elements of the unit circle $\mathbb{R}/2\pi\mathbb{Z}$, as discussed in Section \ref{sec:dual_interpretation}, consecutive mesh points should be approximately uniformly distributed on the unit circle. Consequently, any mesh point $L_i$, together with several adjacent mesh points, should collectively provide uniform coverage of the entire unit circle. Crucially, the mesh points should not cluster around a limited number of positions on the unit circle, as such clustering would result in a loss of information regarding the system dynamics and subsequently degrade the convergence performance of the optimization algorithm.
\end{enumerate}
Methodologies to achieve these conditions are discussed in subsequent subsections. Theoretical interpretation and analysis of these conditions are provided in Section \ref{sec:rationale}.

\subsection{Discretization with Dense Mesh}
As discussed in Section \ref{sec:sundman}, the pseudospectral method converges robustly and accurately with a uniform dense mesh. However, it is still beneficial to satisfy the conditions of monotonicity, near-uniform spacing, and even sampling discussed in Section \ref{sec:mesh_points} to further enhance the performance and accuracy of the optimization algorithm. Specifically, the following two methods improve the performance of the pseudospectral method with a dense mesh:
\begin{enumerate}
    \item Avoid selecting the number of subintervals, denoted by $N$, as an integer multiple of the number of revolutions or as a value approximating such a multiple. When $N$ is close to an integer multiple of the number of revolutions, the mesh points tend to cluster around only a few positions on the unit circle. This clustering results in a loss of information about the system dynamics and degrades the convergence properties of the optimization algorithm.
    \item If the above condition is not satisfied, it is beneficial to incorporate randomness into the mesh points by using the method described in Section \ref{sec:random}. Since the dense mesh points are already well distributed on the unit circle, only a small perturbation is required to ensure full coverage, and the parameters of the method can be adjusted accordingly.
\end{enumerate}
These two approaches correspond to the two methods discussed in Section \ref{sec:sparse} for constructing the mesh points with a sparse mesh. Their effects will be demonstrated through the numerical examples in Section \ref{sec:dense_mesh_solution}.

\subsection{Discretization with Sparse Mesh}\label{sec:sparse}
In certain practical scenarios, such as constructing a comprehensive database for numerous potential orbital transfers, a slight reduction in solution accuracy may be acceptable if it significantly increases the computational efficiency.
We find that the approach outlined in Section \ref{sec:sundman} still converges and provides acceptable solutions even when employing a very sparse mesh (for instance, only a few subintervals per revolution on average, or even fewer than one subinterval per revolution) in combination with the previously discussed low-order collocation. Nonetheless, the mesh points must satisfy the three conditions in Section \ref{sec:mesh_points} to ensure the optimization algorithm's convergence and accuracy.

Two methods are developed to achieve these conditions. The first method is a deterministic approach based on irrational rotation, where a point on the unit circle is advanced by a fixed angle that is an irrational multiple of $2\pi$ at each step. The idea is to choose the number of subintervals $N$ such that the subinterval length cannot be approximated by a rational multiple of $2\pi$ with a small denominator, effectively eliminating the clustering. As a result, the mesh points naturally scatter across the unit circle. The second method incorporates randomness into the positioning of the mesh points. In this approach, random perturbations are generated with a prescribed autocorrelation structure, offering a trade-off between uniform spacing and even sampling. These methods are introduced in the following subsections. Detailed analysis is provided in Section \ref{sec:rationale}, while numerical examples in Section \ref{sec:example} illustrate their performance.

\subsubsection{Strongly Irrational Subinterval Length}\label{sec:irrational}
The first method is to construct a uniform mesh with a specifically chosen subinterval length that eliminates clustering. A uniform discretization naturally satisfies the monotonicity and uniform spacing conditions. However, the even sampling condition may not be met unless the subinterval length meets certain requirements. A uniform discretization with $N$ subintervals yields a mesh spacing of $h = \frac{L_f - L_0}{N}$. The rotation number is defined as
\begin{align}
    \rho = \frac{h}{2 \pi} = \frac{L_f - L_0}{2 \pi N}
\end{align}
To ensure reliable and accurate convergence of the optimization routine, $\rho$ must be sufficiently distant from rational numbers with small denominators. In other words, the fractional part of $\rho$ should not be close to $0$, $1$, $\tfrac{1}{2}$, $\tfrac{1}{3}$, $\tfrac{2}{3}$, or similar fractions with small denominators. Numbers with this property are referred to as strongly irrational in this paper.

Determining whether a real number $\rho$ is close to a rational number is computationally straightforward, and we employ an established method from the theory of Diophantine approximation. For a number $\rho$, its regular continued fraction expansion is defined as
\begin{align}
    \rho = [a_0; a_1, a_2, \dots] = a_0 + \frac{1}{a_1 + \frac{1}{a_2 + \cdots}}
\end{align}
where $a_0$ is the integer part of $\rho$ and $a_1, a_2, \dots$ are the terms of the continued fraction. In the following text, when referring to the continued fraction expansion of $\rho$, it always means the regular continued fraction expansion. A good choice of $N$ will yield small values for the first several terms $a_1, a_2, \dots, a_m$. In contrast, a very large term $a_m$ appearing early (that is, for a small $m$) indicates that $\rho$ is very close to the rational number $[a_0; a_1, a_2, \dots, a_{m-1}]$, which has a small denominator. Such values should therefore be avoided.

As an example, the golden ratio $\phi = \tfrac{1 + \sqrt{5}}{2} \approx 1.618$ has a continued fraction representation of 
\begin{align}
    \phi = [1; 1, 1, 1, \dots] = 1 + \frac{1}{1 + \frac{1}{1 + \cdots}}
\end{align}
and is preferable because it is difficult to approximate by rational numbers~\cite{hardyIntroductionTheoryNumbers2009}. In contrast, $\rho = 1$ is a particularly poor choice. In this case, the mesh points will all differ by multiples of $2\pi$, meaning they actually represent the same point on the unit circle. These two specific values are used in the numerical examples in Section \ref{sec:sparse_mesh_solution} to illustrate the effect of choosing different values of $\rho$.

The regular continued fraction expansion of a real number \( \rho \) is computed through an iterative algorithm that successively extracts integer parts and finds reciprocal of the fractional remainders. Starting with \( x_0 = \rho \), the algorithm iteratively extracts the integer part \( a_k = \lfloor x_k \rfloor \) as the \( k \)-th partial quotient and then computes the next value $x_{k+1} = 1/\left(x_k - a_k\right)$ for the next iteration, provided that \( x_k - a_k \neq 0 \). If \( x_k - a_k = 0 \), then \( \rho \) is a rational number; in that case, the algorithm terminates. The process continues until a stopping criterion is met, such as reaching a maximum number of iterations or achieving a desired level of precision. The resulting sequence \( [a_0; a_1, a_2, \dots, a_m] \) represents the first \( m+1 \) terms of the continued fraction expansion of \( \rho \). This algorithm is computationally efficient and can be implemented using only basic arithmetic operations.

\subsubsection{Randomized Mesh Points}\label{sec:random}
The second method for constructing mesh points introduces randomness into their placement. Given a uniform mesh with $N$ subintervals, we denote the unperturbed uniform mesh points as $L_0, L_1, \dots, L_N=L_f$, where $L_i = L_0 + i h$ for $i = 0, 1, \dots, N$. As noted above, the uniform mesh satisfies the conditions of monotonicity and uniform spacing by construction, but the mesh points may cluster around a few positions on the unit circle if the rotation number $\rho$ is not strongly irrational. The goal is to randomly perturb these mesh points to create a new set $L'_0, L'_1, \dots, L'_N$ that maintains monotonicity while slightly altering the spacing between adjacent mesh points to satisfy the even sampling condition.

To achieve these goals, a procedure based on the autocorrelation of random sequences is developed. Let $d = \min\{h,2\pi\}$. A new set of mesh points is generated by
\begin{equation}
    L'_i = L_i + U_i, \quad i = 1,2,\dots,N-1
\end{equation}
where $U_i$ is drawn from a uniform distribution over $[-d/2,\, d/2]$, with $L'_0 = L_0$ and $L'_N = L_f$. This ensures that the perturbed mesh points remain monotonic while being scattered uniformly on the unit circle, thus fulfilling both the monotonicity and even sampling requirements.

To maintain near-uniform spacing, we additionally require that the perturbations $U_i$ are mutually positively correlated instead of being drawn from independent and identically distributed (i.i.d.) uniform random variables. More specifically, adjacent perturbations $U_i$ and $U_{i+1}$ should have a prescribed Pearson correlation coefficient $r$, which measures the linear correlation between two random variables. The value of $r$ represents a trade-off between the degree of randomness and the uniformity of spacing of the new mesh points. A value of $r$ close to 1 will yield more uniform spacing, resulting in a loss of randomness. Conversely, a value of $r$ close to 0 will yield nearly independent perturbations, producing a more random distribution of mesh points but with high variance in spacing between adjacent points. The value of $r$ should be non-negative for this application. The exact value of $r$ can be adjusted based on the distribution of the original uniform mesh points. If the original mesh points are already well scattered on the unit circle, a higher value of $r$ can be used since only a small perturbation is needed to achieve good coverage of the unit circle. Conversely, if the original mesh points are clustered around a few positions on the unit circle, a lower value of $r$ should be used to ensure that the new mesh points are sufficiently randomized.

The following procedure based on an autoregressive (AR) model is developed to generate the correlated random sequence. The AR model first generates a sequence of random variables that are correlated with a specified Pearson correlation coefficient and normally distributed with mean 0 and variance 1. The normal distribution is then mapped to a uniform distribution using the probability integral transform~\cite{grimmett2001probability}. The procedure to generate correlated random variables uniformly distributed on the unit interval [0,1] is described in the following paragraphs, with pseudocode provided in Algorithm~\ref{alg:random}. Random sequences uniformly distributed in [0,1] can be linearly mapped to the interval $[-d/2,d/2]$ without altering their correlation structure.

First, the target correlation coefficient \(r\) is transformed into the corresponding correlation coefficient \(r_n\) for the Gaussian variables. Because the transformation between the uniform and Gaussian distributions is nonlinear, the correlation coefficient among the uniform variables is not the same as that among the Gaussian variables. The relationship between the two is an established result in the research on Gaussian copulas~\cite{meyerBivariateNormalCopula2013}:
\begin{equation}
    r_n = 2 \sin\left(\frac{\pi r}{6}\right)
\end{equation}
Here, \(r_n\) is the target correlation coefficient for the Gaussian variables.

Next, a Gaussian autoregressive process of order 1 (AR(1)) is generated. The sequence \(\{G_i\}_{i=1}^{N-1}\) is constructed recursively as
\begin{equation}
    G_i = r_n\, G_{i-1} + \sqrt{1 - r_n^2}\,\epsilon_i, \quad \epsilon_i \sim \mathcal{N}(0,1)
\end{equation}
with \(G_1 \sim \mathcal{N}(0,1)\). The symbol \(\mathcal{N}(0,1)\) denotes the standard normal distribution, and \(\epsilon_i\) are i.i.d. standard normal random variables. This model guarantees that each \(G_i\) is a Gaussian random variable with mean 0 and variance 1, and that the correlation between consecutive terms \(G_i\) and \(G_{i+1}\) equals the target value \(r_n\)~\cite{hamilton1994time}.

Finally, the Gaussian variables are transformed into uniform variables using the cumulative distribution function (CDF) of the standard normal distribution
\begin{equation}
    U_i = \Phi(G_i)
\end{equation}
where \(\Phi\) denotes the CDF of the standard normal distribution. This transformation maps the Gaussian variables \(G_i\) onto uniform variables \(U_i\) in the interval \([0,1]\)~\cite{grimmett2001probability}, thereby satisfying the uniform distribution requirement while maintaining the desired correlation structure.

The complete algorithm for generating randomized mesh points \(L'_0, L'_1, \dots, L'_N\) with a prescribed correlation coefficient \(r\) is summarized in Algorithm \ref{alg:random}. The algorithm takes as input the initial and final true longitudes \(L_0\) and \(L_f\), the number of subintervals \(N\), and the desired correlation coefficient \(r\). It outputs the perturbed mesh points \(L'_0, L'_1, \dots, L'_N\) that satisfy the conditions of monotonicity, near-uniform spacing, and even sampling on the unit circle, provided the input parameters are appropriately chosen.

\begin{algorithm}
\caption{Generation of Autocorrelated Random Mesh Points}
\begin{algorithmic}[1]
    \REQUIRE Initial true longitude $L_0$, final true longitude $L_f$, number of subintervals $N$, correlation coefficient $r \in [0,1]$
    \ENSURE Perturbed mesh points $\{L'_i\}_{i=0}^N$ satisfying monotonicity, near-uniform spacing, and even sampling
    
    \STATE \COMMENT{Step 1: Transform target correlation coefficient}
    \STATE $r_n \gets 2 \sin(\pi r / 6)$
    
    \STATE \COMMENT{Step 2: Generate correlated Gaussian random sequence via AR(1) process}
    \STATE $G_1 \sim \mathcal{N}(0, 1)$ \COMMENT{Initialize with standard normal random variable}
    \FOR{$i = 2$ \TO $N-1$}
        \STATE $\epsilon_i \sim \mathcal{N}(0, 1)$ \COMMENT{Independent standard normal random variable}
        \STATE $G_i \gets r_n \cdot G_{i-1} + \sqrt{1 - r_n^2} \cdot \epsilon_i$ \COMMENT{Autoregressive model}
    \ENDFOR
    
    \STATE \COMMENT{Step 3: Transform Gaussian variables to uniform distribution via probability integral transform}
    \FOR{$i = 1$ \TO $N-1$}
        \STATE $U_i \gets \Phi(G_i)$ \COMMENT{$\Phi$ is the standard normal CDF}
    \ENDFOR
    
    \STATE \COMMENT{Step 4: Generate perturbed mesh points with autocorrelated perturbations}
    \STATE $h \gets (L_f - L_0)/N$ \COMMENT{Uniform mesh spacing}
    \STATE $d \gets \min\{h, 2\pi\}$ \COMMENT{Maximum perturbation magnitude}
    \STATE $L'_0 \gets L_0$ \COMMENT{Preserve initial boundary condition}
    \FOR{$i = 1$ \TO $N-1$}
        \STATE $L_i \gets L_0 + ih$ \COMMENT{Unperturbed uniform mesh point}
        \STATE $L'_i \gets L_i + (U_i - 1/2)d$ \COMMENT{Apply scaled perturbation}
    \ENDFOR
    \STATE $L'_N \gets L_f$ \COMMENT{Preserve final boundary condition}
    
    \RETURN $\{L'_i\}_{i=0}^N$
\end{algorithmic}
\label{alg:random}
\end{algorithm}

\section{Rationale of the Methods}\label{sec:rationale}

\subsection{Sundman Transformation and Pseudospectral Methods}
The Sundman transformation is vital to the method. In our experience, the pseudospectral method may fail to converge for multi-revolution transfer problems when using time as the independent variable, even though the problem is well-posed and the mesh is sufficiently dense. This challenge arises because spacecraft dynamics oscillate with respect to the true longitude $L$ with a period of $2\pi$. For a multi-revolution transfer with a large true longitude difference $L_f - L_0$, it is difficult for the optimizer to determine the correct number of revolutions for $L$ in the time domain given these local oscillations. The Sundman transformation addresses this issue by changing the independent variable to $L$, so that each mesh point corresponds to a fixed value of $L$. Although time becomes an additional state variable, the dynamics of the spacecraft do not depend on time, which allows the optimizer to perform the optimization more effectively.
An additional benefit of the Sundman transformation is that a uniform mesh naturally distributes an equal number of subintervals across each orbital revolution, which is advantageous in multi-revolution problems. The number of collocation points within each subinterval is kept minimal to accommodate potential discontinuities in the control inputs, which may exhibit abrupt changes in minimum-fuel orbital transfer scenarios.

\subsection{Discretization with Sparse Mesh}
In this subsection, we explain why the method in Section \ref{sec:method} is effective, even when using fewer than one subinterval per revolution, which might seem insufficient to capture the spacecraft dynamics. As noted in Section \ref{sec:dual_interpretation}, two true longitudes $L_1$ and $L_2$ are defined as equivalent if $L_1 - L_2 = 2\pi k$ for some integer $k$. The effectiveness of the method relies on three key aspects:
\begin{enumerate}
    \item The strongly irrational subinterval length, or the randomized choice of mesh points, ensures that adjacent mesh points are distributed uniformly on the unit circle, effectively sampling the entire circle.
    \item The system, when observed at equivalent true longitudes, exhibits slow evolution and can therefore be integrated accurately with a sparse mesh.
    \item The vector field defined by the dynamical equations (\ref{eq:dotp}--\ref{eq:dotL}) for any two admissible control inputs is nearly commutative, which allows a change in the integration order and the consolidation of control inputs from adjacent revolutions with equivalent true longitudes into a single mesh point.
\end{enumerate}
These aspects are elaborated in detail below. With these properties, the method can be viewed as integrating the slow dynamics using a sparse mesh while simultaneously averaging over the angular variable $L$ using a Monte Carlo approach. The two interpretations of the mesh points discussed in Section \ref{sec:dual_interpretation} are crucial to understanding this process. In their role as points on the real line, the dynamical system is integrated over the range of true longitudes $[L_0, L_f]$. Simultaneously, because the angular variable $L$ is the only fast variable and adjacent mesh points are distributed uniformly on the unit circle by construction, the integration performs numerical averaging over the fast variable $L$ in a Monte Carlo manner. In effect, the system is automatically averaged while integrating over the independent variable $L$.

\begin{figure}
    \centering
    \begin{subfigure}{\textwidth}
        \centering
        \begin{tikzpicture}[scale=0.9]
        \definecolor{blue}{HTML}{397FC7}
        \draw[->, thick] (-0.5,0) -- (10.5,0) node[right] {${L}/{2\pi}$};
        
        \foreach \x in {0,1,...,10} {
            \draw (\x,-0.1) -- (\x,0.1);
            \node[below] at (\x,-0.15) {\x};
        }
        
        \draw[blue, thick, smooth] plot[domain=0:10, samples=50] (\x, {
            1.5 + 0.15*\x
            + 0.2*exp(-0.1*(\x-3)^2)
            - 0.2*exp(-0.1*(\x-7)^2)
        });
        \node[blue] at (9, 3.2) {$\bm{x},\bm{\lambda}$};
        
        \def\theta{70}  
        \foreach \l in {0, 3.2, 6.8, 9.5} {
            \draw[thick, smooth] plot[domain=0:360, samples=50] ({\l+sin(\x)*cos(\theta)}, {cos(\x)+1});
            \foreach \k in {0,45,...,315} {
                \fill ({\l+sin(\k)*cos(\theta)},{cos(\k)+1}) circle [radius=2pt];
            }
        }
        \end{tikzpicture}
        \caption{Traditional averaging method: Mesh points as circle and line elements}
        \label{fig:dual_identity}
    \end{subfigure}
    
    \begin{subfigure}{\textwidth}
        \centering
        \begin{tikzpicture}[scale=0.9]
        \definecolor{blue}{HTML}{397FC7}
        \draw[->, thick] (-0.5,0) -- (10.5,0) node[right] {${L}/{2\pi}$};
        
        \foreach \x in {0,1,...,10} {
            \draw (\x,-0.1) -- (\x,0.1);
            \node[below] at (\x,-0.15) {\x};
        }
        
        \draw[blue, thick, smooth] plot[domain=0:10, samples=50] (\x, {
            1.5 + 0.15*\x
            + 0.2*exp(-0.1*(\x-3)^2)
            - 0.2*exp(-0.1*(\x-7)^2)
        });
        \node[blue] at (9, 3.2) {$\bm{x},\bm{\lambda}$};
        
        \def\theta{70}  

        \draw[thick, smooth] plot[domain=0:10, samples=500] ({\x+cos(360*\x)*cos(\theta)}, {sin(360*\x)+1});
        \foreach \k in {0,1,...,16} {
            \fill ({0.618*\k+cos(360*0.618*\k)*cos(\theta)}, {sin(360*0.618*\k)+1}) circle [radius=2pt];
        }
        \end{tikzpicture}
        \caption{Proposed method: Forward integration with simultaneous circular averaging}
        \label{fig:forward_integration}
        \end{subfigure}
        
        \caption{Illustration of the proposed method compared to traditional averaging methods.}
    \label{fig:sparse_mesh_concept}
\end{figure}
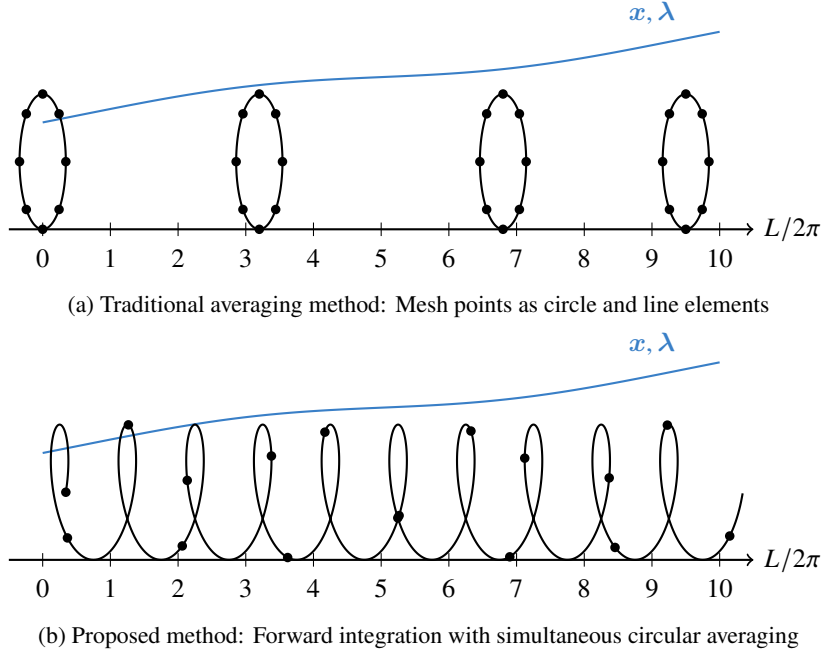

An illustration of this concept is provided in Figure \ref{fig:sparse_mesh_concept}. The upper subfigure \ref{fig:dual_identity} shows the traditional averaging method. As illustrated in the figure, the true longitude $L$ is first integrated over the circle while all other variables are held constant to form an averaged system. The averaged system is then integrated over the real line to obtain the solution. In this case, the mesh points have dual interpretations as points on the real line and as points on the unit circle. An illustration of the proposed method is shown in the lower subfigure \ref{fig:forward_integration}. By carefully selecting the mesh points, the method integrates the system over the true longitude $L$ while simultaneously averaging over the fast variable $L$ on the unit circle. This requires that the mesh points are uniformly distributed both on the real line and on the unit circle. Since all other variables are slow variables, the numerical scheme remains accurate as long as the mesh points are sufficiently uniformly distributed on the unit circle and the system is averaged faster than the secular evolution of the slow variables.

\subsubsection{Strongly Irrational Subinterval Length}
As mentioned in Section \ref{sec:irrational}, the number of subintervals $N$ should be chosen so that the subinterval length cannot be approximated by a rational multiple of $2\pi$ with a small denominator, a property we refer to as ``strongly irrational.'' This choice ensures that the discretized mesh points efficiently sample the entire unit circle in terms of true longitude $L$. An optimal choice of rotation number $\rho$ ensures that each mesh point, together with nearby points, thoroughly covers the unit circle, thereby capturing the full spacecraft dynamics across various true longitudes. Conversely, a suboptimal choice of $\rho$ causes the mesh points to cluster around only a few positions on the unit circle, resulting in significant information loss about the system dynamics and degrading the convergence performance of the optimization algorithm.

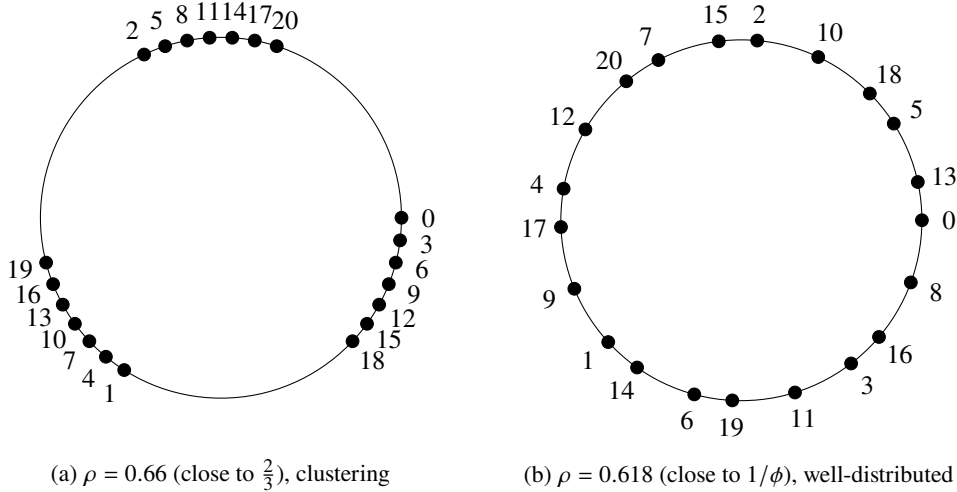
\begin{figure}
    \centering
    \begin{subfigure}{0.45\textwidth}
        \centering
        \begin{tikzpicture}[scale=1.2]
            \draw (0,0) circle (2cm);
            
            \foreach \angle in {0, ..., 20} {
                \filldraw (237.6*\angle:2cm) circle (2pt);
                \node at (237.6*\angle:2.3cm) {\angle};
            }

            \node at (222.48*19:2.3cm) {\phantom{19}};
        \end{tikzpicture}
        \caption{$\rho = 0.66$ (close to $\tfrac{2}{3}$), clustering}
        \label{fig:poorchoice}
    \end{subfigure}
    \begin{subfigure}{0.45\textwidth}
        \centering
        \begin{tikzpicture}[scale=1.2]
            \draw (0,0) circle (2cm);
            
            \foreach \angle in {0, ..., 20} {
                \filldraw (222.48*\angle:2cm) circle (2pt);
                \node at (222.48*\angle:2.3cm) {\angle};
            }
        \end{tikzpicture}
        \caption{$\rho = 0.618$ (close to $1/\phi$), well-distributed}
        \label{fig:goodchoice}
    \end{subfigure}
    \caption{Effect of the rotation number $\rho$ on the uniformity of the discretized mesh points on the unit circle.}
    \label{fig:unitcircle}
\end{figure}

An example of this phenomenon is shown in Figure \ref{fig:unitcircle}. Starting from angle $0$ and successively rotating counterclockwise by a fixed angle of $2\pi\rho$ for each new point, the first 20 points are plotted on the unit circle. The left subfigure \ref{fig:poorchoice} shows a poor choice of $\rho = 0.66$. This value is very close to $\tfrac{2}{3}$, which is a rational number with a small denominator of 3. As a result, the points cluster around only three positions on the unit circle, failing to effectively sample the entire circle. When used as mesh points, this clustering leads to a loss of information about the system dynamics and poor averaging accuracy. In contrast, the right subfigure \ref{fig:goodchoice} shows a good choice of $\rho = 0.618$, which is nearly equal to the reciprocal of the golden ratio. In this case, the mesh points are nearly uniformly distributed on the unit circle, making it an excellent choice for discretization. To avoid clustering, $\rho$ should not be close to any rational number with a small denominator, which can be achieved by checking the regular continued fraction expansion of $\rho$ as described in Section \ref{sec:irrational}.

\subsubsection{Randomized Mesh Points}
Alternatively, randomness can be introduced when generating mesh points. For original mesh points that are uniformly spaced with $h \geq 2\pi$, a uniformly distributed perturbation in the range $[-\pi, \pi]$ can be safely added to each point. Such perturbations preserve the monotonicity of the mesh points. Additionally, all new mesh points (except the first and last ones) will have an equal probability of being positioned anywhere on the unit circle, ensuring uniform distribution.

\begin{figure}
    \centering
    \subcaptionbox{Ranges of random perturbations for mesh points, $\rho = 0.24$.\label{fig:random_range}}[.4\textwidth]{
    \begin{tikzpicture}[scale=1.2]
        \draw (0,0) circle (2cm);
        \def\ang{0.24}
        \definecolor{color1}{HTML}{040676}
        \definecolor{color2}{HTML}{397FC7}
        \definecolor{color3}{HTML}{F1B656}
        
        \tikzset{
            pattern1/.style={pattern=north east lines,pattern color=black},
            pattern2/.style={pattern=grid,pattern color=color2},
            pattern3/.style={pattern=dots,pattern color=color1},
            pattern4/.style={pattern=crosshatch, pattern color=color3}
        }
        
        \foreach \angle/\mypattern in {1/pattern1, 2/pattern2, 3/pattern3, 4/pattern4} {
            \pgfmathparse{\ang*(\angle-0.5)*360}  
            \let\myangles\pgfmathresult       
            \pgfmathparse{\ang*(\angle+0.5)*360}  
            \let\myanglee\pgfmathresult       
            \path[preaction={fill=white}, \mypattern] 
                (\myangles:0) -- (\myangles:2cm) arc (\myangles:\myanglee:2cm) -- cycle;
            \draw[very thick] (\myangles:0) -- (\myangles:2cm) arc (\myangles:\myanglee:2cm) -- cycle;
        }

        \foreach \angle in {0,...,4} {
            \filldraw (\ang*360*\angle:2cm) circle (2pt);
            \node at (\ang*360*\angle:2.3cm) {\angle};
        }
    \end{tikzpicture}
    }
    \hspace{0.05\textwidth}
    \subcaptionbox{Example of autocorrelated uniform sequences. The sequences become less volatile as $r$ increases.\label{fig:autocorrelated_example}}
    {
        \centering
        \includegraphics[width=0.45\textwidth]{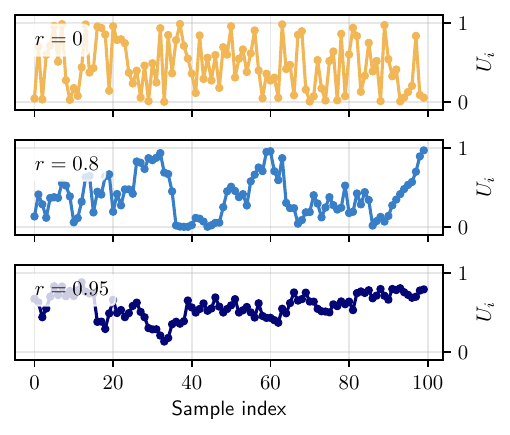}
    }
    \caption{Illustration of the core concepts underlying the construction of randomized mesh points.}
\end{figure}
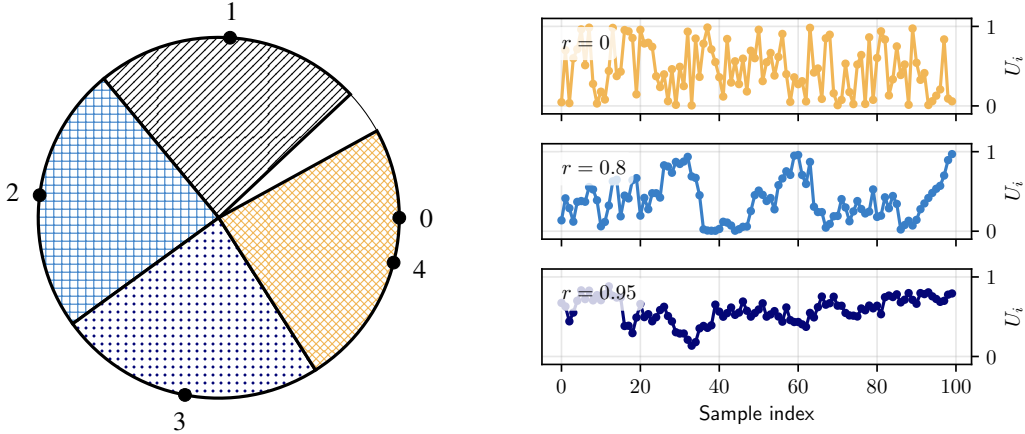

On the other hand, if the original mesh points are uniformly spaced with $h < 2\pi$, the perturbation range must be limited to $[-h/2, h/2]$ to ensure that the new mesh points remain monotonic. In this case, several consecutive mesh points together cover the unit circle. Figure~\ref{fig:random_range} illustrates an example with $\rho = 0.24$, which is close to $\tfrac{1}{4}$ and not strongly irrational. In the figure, the solid points represent the unperturbed uniform mesh points $L_i$. Since $\rho$ is near $\tfrac{1}{4}$, these uniformly spaced mesh points cluster around four locations on the unit circle. To overcome this clustering, random perturbations are applied. The resulting perturbed mesh points $L'_1, L'_2, L'_3, L'_4$ are selected within the four differently colored sectors shown in the figure. These sectors collectively cover the entire unit circle while ensuring the monotonicity of the mesh points is preserved.

The main drawback of randomized mesh points is that the perturbed mesh points no longer have uniform spacing. In pseudospectral methods, the integration weights must be adjusted accordingly. This unevenness in the integration weights results in unequal contributions when averaging the fast variable $L$, thereby increasing the overall numerical error. To address this, the random perturbations described in Section \ref{sec:random} are autocorrelated, while each perturbation remains uniformly distributed within its range. An illustration of the autocorrelated uniform sequences is shown in Figure \ref{fig:autocorrelated_example}. The first 100 randomly generated values of $U_i$ are shown for correlation coefficients $r=0$, $0.8$, and $0.95$. As $r$ increases, the sequences become less volatile. With this approach, if the mesh point $L_i$ is perturbed to $L'_i = L_i + U_i$, the next perturbation $U_{i+1}$ will tend to be similar to $U_i$, thereby reducing the deviation of $L'_{i+1} - L'_i$ from the original spacing $h$. A larger $r$ encourages more uniform spacing between adjacent perturbed mesh points. However, if $r$ is too large, the perturbations become overly homogeneous, resulting in insufficient randomness. Therefore, a balance must be struck between randomness and uniformity in the spacing of the perturbed mesh points.

\subsubsection{Slow Dynamics at Equivalent True Longitudes}\label{sec:slow_dynamics}
The state, costate, and control variables follow a similar pattern over adjacent revolutions, varying only slightly at equivalent true longitude values $L$ and $L \pm 2\pi$. From the perspective of averaging methods, at true longitudes $L$ and $L \pm 2\pi$, the state and costate variables, except for the true longitude $L$ itself, exhibit only secular variations that are slow compared to the rapid oscillations of $L$. Specifically, in a well-posed multi-revolution problem with low-thrust control, aside from the control inputs $\bd{u}$, the costate variable $\lambda_L$ corresponding to the true longitude $L$ is very small. The value of $\lambda_L$ measures the sensitivity of the objective function to $L$, since shifting the spacecraft's phase only slightly changes the cost by altering the orbit's semi-major axis. Consequently, all terms in the Hamiltonian are small, and $L$ is the only rapidly varying variable. Therefore, the state variables (excluding $L$) $[p, f, g, h, k, m]^\rmt$, and the costate variables (including $\lambda_L$) $[\lambda_p, \lambda_f, \lambda_g, \lambda_h, \lambda_k, \lambda_L, \lambda_m]^\rmt$, vary more slowly than $L$, remaining nearly constant between adjacent revolutions.

As required in Equation~\eqref{eq:objective}, the objective depends solely on the magnitude of the control inputs. Hence, the optimal control variable $\bd{u}$ can be determined by first finding the optimal direction and then the corresponding magnitude. The direction of the optimal control inputs changes only slightly between adjacent revolutions at equivalent true longitudes $L$ and $L \pm 2\pi$. The magnitude may vary continuously with respect to the state and costate variables or exhibit abrupt jumps in bang-bang control cases. However, because the throttle function depends continuously on these variables, the locations of these jumps change slowly between adjacent revolutions. Consequently, only a few nodes are affected, and the overall impact on the convergence of the optimization algorithm remains small, as demonstrated by the numerical experiments in Section~\ref{sec:sparse_mesh_solution}. We explain these two aspects in more detail in the following paragraphs.

Under the assumption that the objective function $F$ depends only on the magnitude of the control inputs, the Hamiltonian of the system can be written as
\begin{equation}
    H = F + \bd{\lambda}^\rmt \dot{\bd{x}} = H_0(\bd{x}, \bd{\lambda}, \|\bd{u}\|) + \bd{H}_1^\rmt(\bd{x}, \bd{\lambda})\bd{u}
\end{equation}
where 
\begin{equation}
    H_0 = F(\|\bd{u}\|) + \lambda_L \sqrt{\mu p} \left(\frac{w}{p}\right)^2 - \lambda_m \frac{\|\bd{u}\|}{I_\text{sp} g_0}
\end{equation}
and $\bd{H}_1^\rmt\bd{u}$ contains all the remaining terms in the Hamiltonian. To minimize $H$ with respect to $\bd{u}$, as required by Pontryagin's maximum principle, the direction of the optimal control vector $\bd{u}$ is always aligned with $-\bd{H}_1$. Therefore, the direction of $\bd{u}$ varies continuously with respect to the state variables $\bd{x}$ and the costate variables $\bd{\lambda}$.

Regarding thrust magnitude, its continuity with respect to $\bd{x}$ and $\bd{\lambda}$ depends on whether the second derivative of $H$ with respect to $\|\bd{u}\|$ vanishes. A common case in practice is when $H$ depends linearly on $\|\bd{u}\|$. In this scenario, $\|\bd{u}\|$ may exhibit abrupt changes at certain points. However, because the coefficients of $\|\bd{u}\|$ (viewed as the throttle function) depend continuously on $\bd{x}$ and $\bd{\lambda}$, the burning and coasting arcs vary only slightly between adjacent revolutions. Consequently, the control input at most mesh points $L$ remains representative of adjacent revolutions with equivalent true longitudes $L \pm 2\pi$. A concrete example for the bang-bang control case is shown in Section~\ref{sec:dense_mesh_solution} and Figure~\ref{fig:thrust} of the numerical example. 

\subsubsection{Near Commutativity of the Vector Fields}\label{sec:commutativity}
In pseudospectral methods, the dynamics and objective function integration are discretized at a finite number of mesh and collocation points. As shown in the analysis above, the optimal control input value at one mesh node $L_i$ is very close to its value at $L_i \pm 2k\pi$ for $k=0,1,\dots,n$, with $n$ being a small integer. Therefore, real mesh points can be ``unfolded" into $2n$ virtual ones, $L_i \pm 2k\pi,\; k=1,\dots,n$. Optimal control inputs at the virtual mesh points are assumed to be the same as those at the real mesh points, with integration weights split accordingly. The choice of mesh points ensures that enough virtual points are distributed across each revolution. Integrating these unfolded points in ascending order of $L$ provides a good approximation of the continuous problem. Using the real mesh point order effectively reorders the integration over these virtual points.

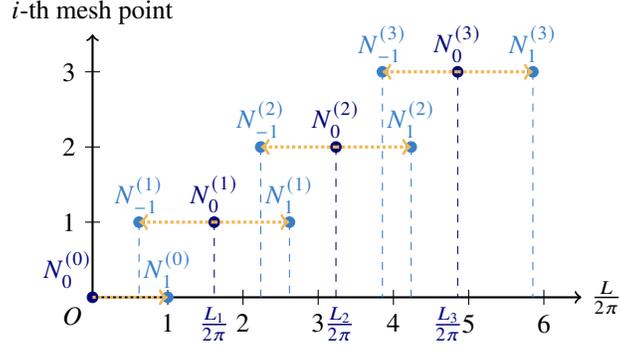
\begin{figure}
    \centering
\begin{tikzpicture}
    \def\xmax{6} 
    \def\ymax{3} 
    \def\phi{1.618} 

    \definecolor{red}{HTML}{040676}
    \definecolor{blue}{HTML}{397FC7}
    \definecolor{grey}{HTML}{F1B656}

    \draw[->, thick] (0,0) -- (\xmax+0.5,0) node[right]{$\frac{L}{2\pi}$}; 
    \draw[->, thick] (0,0) -- (0,\ymax+0.5) node[above]{$i$-th mesh point}; 
    
    \node at (0,0) [below left]{$O$};
    
    \foreach \x in {1,...,\xmax} {
        \draw (\x,0.1) -- (\x,-0.1) node[below]{$\x$};
    }
    \foreach \y in {1,...,\ymax} {
        \draw (0.1,\y) -- (-0.1,\y) node[left]{$\y$};
    }

    \filldraw[red] (0,0) circle (2pt); 
    \filldraw[red] (\phi, 1) circle (2pt); 
    \filldraw[red] (2*\phi, 2) circle (2pt); 
    \filldraw[red] (3*\phi, 3) circle (2pt); 

    \node[red, above] at (-0.33,0) {$N^{(0)}_0$};
    \node[red, above] at (\phi, 1)  {$N^{(1)}_0$};
    \node[red, above] at (2*\phi, 2)  {$N^{(2)}_0$};
    \node[red, above] at (3*\phi, 3) {$N^{(3)}_0$};

    \draw[red, dashed] (\phi, 1) -- (\phi, 0) node[below]{$\frac{L_1}{2\pi}$};
    \draw[red, dashed] (2*\phi, 2) -- (2*\phi, 0);
    \node[red, below] at (2*\phi+0.05, 0) {$\frac{L_2}{2\pi}$};
    \draw[red, dashed] (3*\phi, 3) -- (3*\phi, 0);
    \node[red, below] at (3*\phi-0.13, 0) {$\frac{L_3}{2\pi}$};
    
    \filldraw[blue] (1, 0) circle (2pt); 
    \filldraw[blue] (\phi-1, 1) circle (2pt); 
    \filldraw[blue] (\phi+1, 1) circle (2pt); 
    \filldraw[blue] (2*\phi-1, 2) circle (2pt); 
    \filldraw[blue] (2*\phi+1, 2) circle (2pt); 
    \filldraw[blue] (3*\phi-1, 3) circle (2pt); 
    \filldraw[blue] (3*\phi+1, 3) circle (2pt); 

    \node[blue, above] at (1, 0) {$N^{(0)}_1$};
    \node[blue, above] at (\phi-1, 1) {$N^{(1)}_{-1}$};
    \node[blue, above] at (\phi+1, 1) {$N^{(1)}_{1}$};
    \node[blue, above] at (2*\phi-1, 2) {$N^{(2)}_{-1}$};
    \node[blue, above] at (2*\phi+1, 2) {$N^{(2)}_{1}$};
    \node[blue, above] at (3*\phi-1, 3) {$N^{(3)}_{-1}$};
    \node[blue, above] at (3*\phi+1, 3) {$N^{(3)}_{1}$};

    \draw[blue, dashed] (\phi-1, 1) -- (\phi-1, 0);
    \draw[blue, dashed] (\phi+1, 1) -- (\phi+1, 0);
    \draw[blue, dashed] (2*\phi-1, 2) -- (2*\phi-1, 0);
    \draw[blue, dashed] (2*\phi+1, 2) -- (2*\phi+1, 0);
    \draw[blue, dashed] (3*\phi-1, 3) -- (3*\phi-1, 0);
    \draw[blue, dashed] (3*\phi+1, 3) -- (3*\phi+1, 0);

    \draw[->, grey, densely dotted, very thick] (0, 0) -- (1, 0);
    \draw[->, grey, densely dotted, very thick] (\phi, 1) -- (\phi-1, 1);
    \draw[->, grey, densely dotted, very thick] (\phi, 1) -- (\phi+1, 1);
    \draw[->, grey, densely dotted, very thick] (2*\phi, 2) -- (2*\phi-1, 2);
    \draw[->, grey, densely dotted, very thick] (2*\phi, 2) -- (2*\phi+1, 2);
    \draw[->, grey, densely dotted, very thick] (3*\phi, 3) -- (3*\phi-1, 3);
    \draw[->, grey, densely dotted, very thick] (3*\phi, 3) -- (3*\phi+1, 3);
\end{tikzpicture}
\caption{Unfolding and reordering of the mesh points. }
\label{fig:reordering}
\end{figure}

An example of this reordering is shown in Figure \ref{fig:reordering}. The figure illustrates the case of $\rho = 1.618$ with the strongly irrational rotation method. In the figure, the $i$-th real mesh point is denoted by $N^{(i)}_0$ and is plotted in deep blue. Each real mesh point is unfolded into two virtual mesh points, with a difference in $L$ of $2\pi$. The virtual mesh points, $N^{(i)}_{-1}$ and $N^{(i)}_{1}$, are plotted in light blue. The integration order for the unfolded mesh should be $N^{(0)}_0, N^{(1)}_{-1}, N^{(0)}_1, N^{(1)}_0, N^{(2)}_{-1},\allowbreak N^{(1)}_1, N^{(2)}_0, N^{(3)}_{-1}, N^{(2)}_1, N^{(3)}_0, N^{(3)}_{1}$, following the ascending order of their coordinates in the horizontal axis representing the true longitude $L$. 
However, with only the real mesh points available, the integration order is effectively $[N^{(0)}_0, N^{(0)}_1], [N^{(1)}_{-1}, N^{(1)}_0, N^{(1)}_{1}], [N^{(2)}_{-1}, N^{(2)}_0,\allowbreak N^{(2)}_{1}],\allowbreak [N^{(3)}_{-1},\allowbreak N^{(3)}_0, N^{(3)}_{1}]$, which follows the ascending order of coordinates in the vertical axis. The square brackets indicate that the virtual mesh points are integrated together as a single real mesh point. A concrete example of this reordering is presented and analyzed in Section~\ref{sec:strongly_irrational_solution} and Figure~\ref{fig:compare_states} of the numerical example.

A system of dynamical equations defines a vector field by specifying a tangent vector at each point in the phase space, and the solutions to the system generate the flow of the vector field, describing the evolution of states over time. To validate the method and justify the reordering of the integration, another core mechanism is the near commutativity of any two vector fields generated by the dynamical equations when different control variables $\bd{u}$ are applied. Since $L$ is the fast variable, a small change in $L$ always corresponds to a small change in the time variable $t$, and vice versa. Therefore, it is equivalent to work with the dynamical equations in either the time domain or the true longitude domain because they inherently describe the same system. The following analysis will be conducted in the time domain before the Sundman transformation for simplicity.

The objective is to demonstrate that for any initial state $\bm{x}_0$ and two admissible control inputs $\bm{u}_1$ and $\bm{u}_2$ applied over sufficiently small time durations $s$ and $t$, respectively, the resulting system states exhibit negligible variation regardless of the order in which the controls are applied. This can be expressed as
\begin{align}
    \phi^{\bm{u}_1}_{s} \circ \phi^{\bm{u}_2}_{t}(\bm{x}_0) \approx \phi^{\bm{u}_2}_{t} \circ \phi^{\bm{u}_1}_{s}(\bm{x}_0),
\end{align}
where $\phi^{\bm{u}}_t(\bm{x}_0)$ represents the state evolution under control input $\bm{u}$ for time $t$ starting from $\bm{x}_0$, and $\circ$ denotes function composition.  

This result follows intuitively because the vector fields governing the dynamics depend weakly on the control inputs due to their small magnitudes. To quantify this, we examine the Lie bracket $[\bm{P},\bm{Q}]$ of the corresponding vector fields $\bm{P}$ and $\bm{Q}$. The Lie bracket, defined as $[\bm{P},\bm{Q}] = \bm{P}\bm{Q} - \bm{Q}\bm{P}$, measures the non-commutativity of $\bm{P}$ and $\bm{Q}$. In local coordinates, if $\bm{P} = \sum_i P^i \partial_i$ and $\bm{Q} = \sum_j Q^j \partial_j$, then
\begin{align}
    \bm{P}(\bm{Q}f) = \sum_{i,j} P^i (\partial_i Q^j) \partial_j f + P^i Q^j \partial_i \partial_j f.
\end{align}
The Lie bracket cancels the symmetric second-order terms $\partial_i \partial_j f$, leaving only first-order derivatives and ensuring $[\bm{P},\bm{Q}]$ remains a vector field~\cite{leeIntroductionSmoothManifolds2012}.  

The non-commutativity of the flows $\phi^{\bm{P}}_s$ and $\phi^{\bm{Q}}_t$ can be estimated as~\cite{leeIntroductionSmoothManifolds2012}
\begin{equation}
    \phi^{\bm{P}}_s \circ \phi^{\bm{Q}}_t(\bm{x}) - \phi^{\bm{Q}}_t \circ \phi^{\bm{P}}_s(\bm{x}) = st [\bm{P},\bm{Q}](\bm{x}) + \text{higher-order terms}
\end{equation}
When $\|[\bm{P},\bm{Q}]\|$ is of order $u_{\text{max}}$ (the maximum control magnitude, assumed small), the leading term $st [\bm{P},\bm{Q}]$ becomes negligible. This confirms that the flows are nearly commutative, justifying the initial approximation.

The vector fields governing the system dynamics are expressed in local coordinates as linear combinations of partial derivatives $\partial/\partial \bm{x}_i$ with smooth coefficient functions. The objective function $F(\|\bm{u}\|)$ is integrated into the vector field formulation alongside the orbital dynamics. For a control input $\bm{u}$, the associated vector field $\bm{X}^{\bm{u}}$ is explicitly defined as
\begin{equation}
    \begin{aligned}
    \bd{X}^{\bd{u}} &= \frac{2 p a_t}{w} \sqrt{\frac {p}{\mu}}  \frac{\partial}{\partial p}
        + \sqrt{\frac{p}{\mu}} \left\{  a_r \sin L +\frac{\left[(w+1)\cos L+f\right]a_t}{w} - \frac{g(h\sin L-k\cos L)a_n}{w}\right\}  \frac{\partial}{\partial f} \\
        &\quad+ \sqrt{\frac{p}{\mu}} \left\{  -a_r \cos L +\frac{\left[(w+1)\sin L+g\right]a_t}{w} + \frac{f(h\sin L-k\cos L)a_n}{w}\right\}  \frac{\partial}{\partial g} \\
        &\quad+ \sqrt{\frac{p}{\mu}} \frac{s^2a_n}{2w}\cos L  \frac{\partial}{\partial h} 
        +\sqrt{\frac{p}{\mu}} \frac{s^2a_n}{2w}\sin L  \frac{\partial}{\partial k} + \left[\sqrt{\mu p} \left(\frac{w}{p}\right)^2+ \sqrt{\frac{p}{\mu}} \frac{(h\sin L-k\cos L)a_n}{w}\right]  \frac{\partial}{\partial L} \\
        &\quad-\frac {\|\bd{u}\|}{I_\text{sp} g_0}  \frac{\partial}{\partial m}
        + F(\|\bd{u}\|)  \frac{\partial}{\partial J}
    \end{aligned}
\end{equation}

To quantify the non-commutativity of control actions, we compute the Lie bracket of two control-generated vector fields $\bm{X}^{\bm{u}_1}$ and $\bm{X}^{\bm{u}_2}$
\begin{equation}
    \begin{aligned}
    \left[\bm{X}^{\bm{u}_1}, \bm{X}^{\bm{u}_2}\right] &= \frac{2 \mu (g \cos L - f \sin L) w^3}{p^3} \frac{\partial}{\partial L} + O\left(u_\text{max}\right) \\
    &\quad - \frac{2 \mu (g \cos L - f \sin L) w^3}{p^3} \frac{\partial}{\partial L} - O\left(u_\text{max}\right) \\
    &= O\left(u_\text{max}\right).
    \end{aligned}
\end{equation}
The dominant first-order terms cancel exactly due to their opposing signs, leaving only higher-order residuals proportional to $u_\text{max}$. Under the assumption of small control magnitudes, these residuals become negligible. This result aligns with the desired property of flow commutativity: the Lie bracket's $O(u_\text{max})$ norm confirms that the state variation induced by reordering $\bm{u}_1$ and $\bm{u}_2$ is insignificant over small time intervals. Consequently, the integration sequence in the proposed method can be interchanged while still preserving the accuracy of the trajectory optimization framework.

\section{Numerical Results}\label{sec:example}
In this section, we illustrate the performance of the proposed method using a minimum-fuel rendezvous problem that is highly elliptical, non-planar, includes the $J_2$ perturbation, and involves 250 revolutions with bang-bang control. This problem is considered challenging to solve; however, we show that the proposed method can handle it efficiently and accurately.

All numerical experiments are conducted using our self-developed software, pockit, a Python package for optimal control pseudospectral methods. The source code for this package is freely available online\footnote{\url{https://github.com/zouyilin2000/pockit}}. We employ the integral Legendre-Gauss-Lobatto (LGL) method with two collocation points per subinterval. The resulting optimization problem is solved using IPOPT~\cite{wachterImplementationInteriorpointFilter2006} with MUMPS~\cite{amestoyFullyAsynchronousMultifrontal2001, amestoyPerformanceScalabilityBlock2019} as the linear solver and default settings. The tests are performed on an Apple M4 chip. We set the initial guess for the state variables at each mesh point by linearly interpolating between the initial and final boundary conditions, except for the mass, which remains fixed at its initial value since its final value is free. The initial guess for each component of the control inputs is set to nearly zero (\num{e-10}). A unit normalization is used, where the length unit is the Earth's radius $R_e$ of \qty{6378.1363}{\kilo\meter}, the time unit is the transfer duration of \qty{190}{days}, and the mass unit is the initial spacecraft mass of \qty{2000}{\kilo\gram}.

\subsection{Problem Setup}
We consider a low-thrust, minimum-fuel orbital rendezvous problem from a geostationary transfer orbit (GTO) to a geostationary orbit (GEO). The spacecraft is equipped with a low-thrust propulsion system that has a specific impulse of \qty{3000}{\second} and a maximum thrust of \qty{0.5}{\newton}, with an initial mass of \qty{2000}{\kilo\gram}. The initial and final states of the spacecraft are provided in MEE in Table \ref{tab:initial_final_states}, using the same values as in~\cite{wuWarmStartMultihomotopicOptimization2020}. In the table, the semi-latus rectum $p$ is given in units of the Earth's radius $R_e$, which is \qty{6378.1363}{\kilo\meter}. Note that in orbital rendezvous problems, the number of revolutions for the optimal trajectory is typically unknown. In general, one must optimize over the number of revolutions, which is an integer variable, and adjust the boundary condition $L_f$ accordingly. Estimating and optimizing the number of revolutions is beyond the scope of this paper.

\begin{table}
    \centering
    \caption{Initial and final states of the spacecraft in MEE.}
    \begin{tabular}{cccccccc}
        \toprule
        $p$ ($R_e$) & $f$ & $g$ & $h$ & $k$ & $L$ (\unit{\radian}) & $m$ (\unit{\kilo\gram}) & $t$ (\unit{day}) \\
        \midrule
        \num{1.7787} & \num{-0.1144} & \num{0.722} & \num{-0.0376} & \num{0.2371} & \num{4.89} & \num{2000} & \num{0} \\
        \num{6.6107} & \num{1.7637e-7} & \num{-1.39e-6} & \num{0} & \num{0} & \num{1575.635} & Free & \num{190} \\
        \bottomrule
    \end{tabular}
    \label{tab:initial_final_states}
\end{table}

In addition to the orbital dynamics described in Section~\ref{sec:problem}, the $J_2$ perturbation is incorporated to account for the Earth's oblateness. The $J_2$ perturbation represents a second-order zonal harmonic term in Earth's gravitational potential and produces additional accelerations in the radial, transverse, and normal directions. Specifically, as given in~\cite{walkerSetModifiedEquinoctial1985}
\begin{align}
    a^{\left(J_2\right)}_r &= -\frac{3 \mu J_2 R_e^2}{2 r^4} \left[ 1 - \frac{12 (h \sin L - k \cos L)^2}{(1 + h^2 + k^2)^2} \right] \\
    a^{\left(J_2\right)}_t &= -\frac{12 \mu J_2 R_e^2}{r^4} \left[ \frac{(h \sin L - k \cos L)(h \cos L + k \sin L)}{(1 + h^2 + k^2)^2} \right] \\
    a^{\left(J_2\right)}_n &= -\frac{6 \mu J_2 R_e^2}{r^4} \left[ \frac{(1 - h^2 - k^2)(h \sin L - k \cos L)}{(1 + h^2 + k^2)^2} \right]
\end{align}
For the Earth, the $J_2$ coefficient is set to \num{1.08262668e-3} and $R_e$ denotes Earth's radius.

\subsection{Dense Mesh Solution}\label{sec:dense_mesh_solution}
We first solve the problem using a dense mesh with 2222, 2500, 6666, and 10000 subintervals, corresponding to 8.888, 10, 26.664, and 40 subintervals per revolution, respectively. The numbers 2222 and 6666 are deliberately chosen to avoid being integer multiples of the number of revolutions, while 2500 and 10000 are integer multiples of the number of revolutions. The results are shown in Table \ref{tab:dense_mesh}. In the table, the subscript $c$ denotes the solution with two-body dynamics without perturbation, and the subscript $J_2$ denotes the solution with the $J_2$ perturbation. The time indicates the total computational time of the optimization process, and the objective represents the total fuel consumption. With uniform meshes, the objective for the two-body dynamics case with 2222 and 6666 subintervals matches the value $\qty{135.65}{\kilogram}$ reported in the literature\cite{wuWarmStartMultihomotopicOptimization2020}, while those with 2500 and 10000 exhibit a small difference. The randomization method introduced in Section \ref{sec:random} is further applied to the 2500- and 10000-subinterval cases. Since the mesh is already dense, a slight perturbation is enough to achieve good coverage of the unit circle, so the correlation coefficient $r$ is set to 0.99. The results show that the randomization method reduces both the computational time and the error in the objective function in the two-body dynamics case compared to unperturbed uniform mesh.

\begin{figure}
    \centering
    \includegraphics[width=0.8\textwidth]{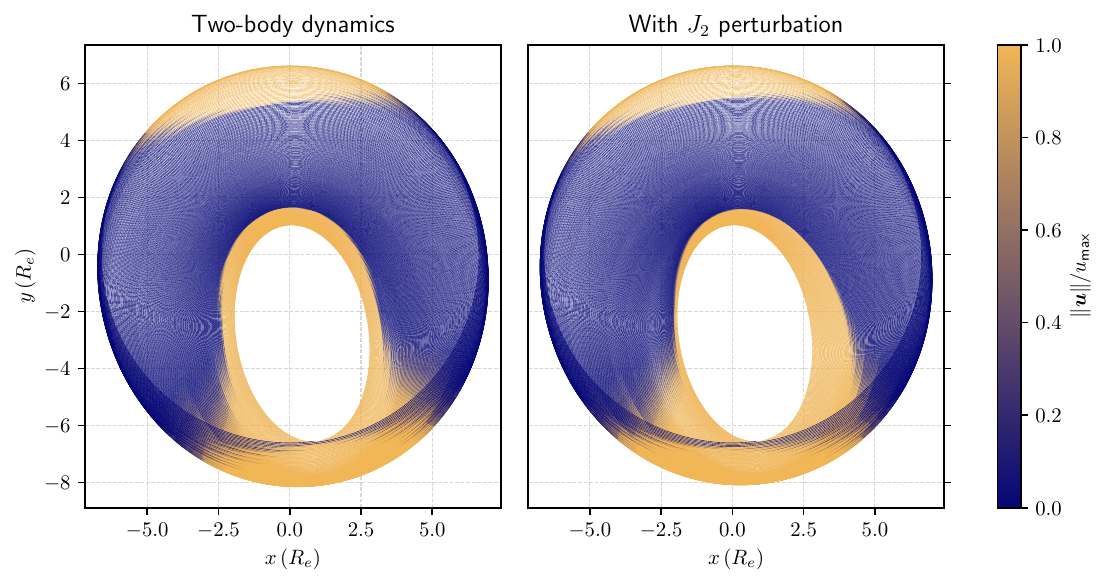}
    \caption{The optimal trajectory obtained using 6666 subintervals is projected onto the $x$-$y$ plane. Orange lines represent burning arcs, and blue lines represent coasting arcs.}
    \label{fig:trajectory}
\end{figure}

\begin{figure}
    \centering
    \includegraphics[width=0.8\textwidth]{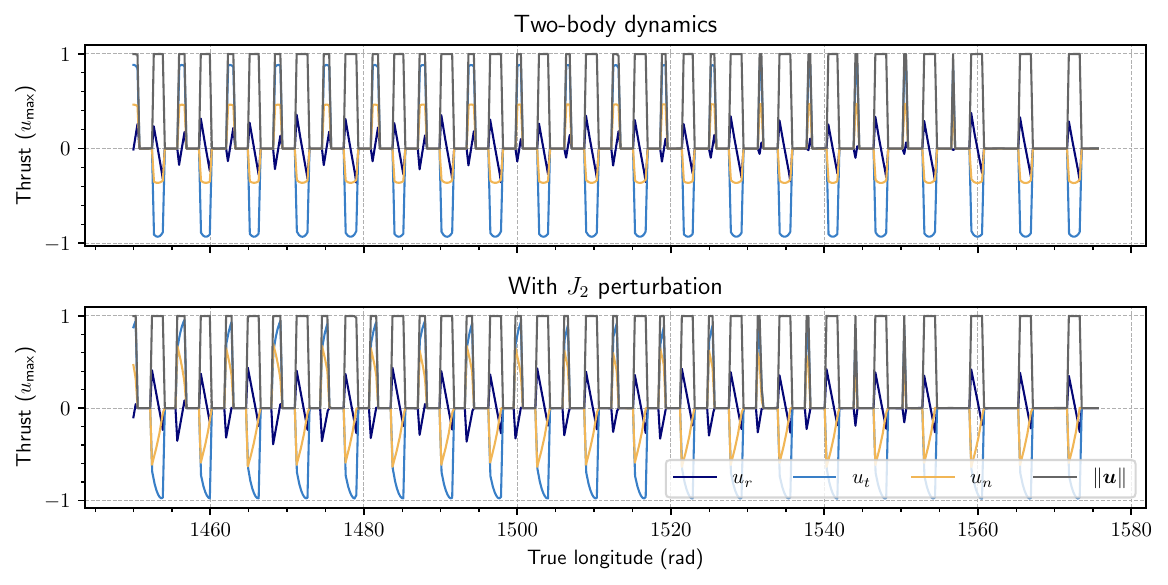}
    \caption{Optimal control inputs obtained using 6666 subintervals during the final 20 revolutions.}
    \label{fig:thrust}
\end{figure}

In the following text, we use the solutions obtained with 6666 subintervals as the reference for evaluating the optimal objectives of the sparse mesh solutions, for both cases with and without the $J_2$ perturbation. The optimal trajectory obtained with 6666 subintervals is shown in Figure~\ref{fig:trajectory}. Because the transfer trajectory is inclined, some overlap appears near the apogees in the figure. Figure~\ref{fig:thrust} displays the thrust direction and magnitude during the last 20 revolutions. The thrust magnitude exhibits a bang-bang pattern, as expected for minimum-fuel problems. The thrust follows a similar pattern across adjacent revolutions, which aligns with the analysis in Section~\ref{sec:slow_dynamics}.

\begin{table}
    \centering
    \caption{Results for the dense mesh solution.}
    \begin{tabular}{cccccc}
        \toprule
        Number of subintervals & Mesh Type & $\text{Time}_c$ (\unit{\second}) & $\text{Objective}_c$ (\unit{\kilo\gram}) & $\text{Time}_{J_2}$ (\unit{\second}) & $\text{Objective}_{J_2}$ (\unit{\kilo\gram}) \\
        \midrule
        \num{2222} & Uniform & \num{13.721} & \num{135.654402} & \num{16.021} & \num{140.308377} \\
        \num{2500} & Uniform& \num{19.352} & \num{137.708628} & \num{15.326}  & \num{141.390220} \\
        \num{2500} & Randomized & \num{16.649} & \num{135.450706} & \num{14.434}  & \num{140.243720} \\
        \num{6666} & Uniform& \num{43.319} & \num{135.655953} & \num{48.923} & \num{140.305407} \\
        \num{10000} & Uniform& \num{75.918} & \num{135.742796} & \num{187.256} & \num{140.300731} \\
        \num{10000} & Randomized & \num{70.780} & \num{135.654954} & \num{97.584}  & \num{140.311068} \\
        \bottomrule
    \end{tabular}
    \label{tab:dense_mesh}
\end{table}

\subsection{Sparse Mesh Solution}\label{sec:sparse_mesh_solution}
In this section, we evaluate the effectiveness of the proposed methods using a sparse mesh. Employing a sparse mesh substantially reduces the computational time of the optimization process. As the subsequent results demonstrate, the optimization procedure can typically be completed within a few seconds, which is a significant improvement over the dense mesh approach. Furthermore, our findings show that by implementing the methods introduced in Section \ref{sec:sparse}, namely, selecting strongly irrational subinterval lengths or applying random perturbations with autocorrelation, the solution accuracy remains highly satisfactory and suitable for practical applications.

\subsubsection{Strongly Irrational Subinterval Length}\label{sec:strongly_irrational_solution}
To test the strongly irrational subinterval length method, the problem is discretized with approximately 50, 250, and 410 subintervals, corresponding to 0.2, 1, and 1.64 subintervals per revolution, respectively. The results are shown in Tables \ref{tab:sparse_mesh_50}, \ref{tab:sparse_mesh_250}, and \ref{tab:sparse_mesh_410}. In these tables, $N$ is the number of subintervals, $\rho$ is the rotation number defined as the subinterval length divided by $2\pi$, the continued fraction of $\rho$ is the first few terms of its expansion with the integer part $a_0$ omitted, $T$ is the total computational time, $J$ is the obtained objective, and $E$ is the relative error of the objective with respect to the reference result. The subscript $c$ denotes the solution with the two-body model (no perturbation), and the subscript $J_2$ denotes the solution with the $J_2$ perturbation as in the dense mesh results.

\begin{table}
    \centering
    \caption{Results for the sparse mesh solution using approximately 50 subintervals.}
    \begin{tabular}{ccccccccc}
        \toprule
        $N$ & $\rho$ & continued fraction of $\rho$ & $T_c$ (\unit{\second}) & $J_c$ (\unit{\kilo\gram}) & $E_c$ (\unit{\percent}) & $T_{J_2}$ (\unit{\second}) & $J_{J_2}$ (\unit{\kilo\gram}) & $E_{J_2}$ (\unit{\percent}) \\
        \midrule
        \num{40} & \num{6.250} & [4, 305, 2, 1, 5] & \num{1.135} & \num{121.096} & \num{10.733} & \num{1.181} & \num{131.931} & \num{5.969}\\
        \num{41} & \num{6.097} & [10, 3, 1, 2, 4] & \num{1.121} & \num{135.224} & \num{0.318} & \num{0.802} & \num{140.353} & \num{0.034}\\
        \num{42} & \num{5.952} & [1, 19, 1, 10, 1] & \num{1.161} & \num{139.345} & \num{2.719} & \num{0.892} & \num{143.287} & \num{2.125}\\
        \num{43} & \num{5.814} & [1, 4, 2, 1, 2] & \num{1.427} & \num{136.225} & \num{0.419} & \num{1.031} & \num{139.806} & \num{0.356}\\
        \num{44} & \num{5.682} & [1, 2, 7, 10, 1] & \num{1.580} & \num{139.922} & \num{3.145} & \num{0.939} & \num{140.512} & \num{0.147}\\
        \num{45} & \num{5.555} & [1, 1, 4, 67, 1] & \num{1.106} & \num{137.357} & \num{1.254} & \num{1.205} & \num{142.764} & \num{1.753}\\
        \num{46} & \num{5.435} & [2, 3, 3, 10, 3] & \num{1.208} & \num{136.132} & \num{0.351} & \num{0.916} & \num{139.174} & \num{0.807}\\
        \num{47} & \num{5.319} & [3, 7, 2, 2, 7] & \num{1.462} & \num{140.027} & \num{3.223} & \num{1.108} & \num{145.503} & \num{3.704}\\
        \num{48} & \num{5.208} & [4, 1, 4, 9, 1] & \num{1.040} & \num{137.124} & \num{1.082} & \num{1.366} & \num{140.141} & \num{0.117}\\
        \num{49} & \num{5.102} & [9, 1, 4, 2, 3] & \num{1.410} & \num{134.634} & \num{0.754} & \num{1.015} & \num{140.606} & \num{0.214}\\
        \num{50} & \num{5.000} & [1, 6111, 20, 1, 5] & Failed & - & - & Failed & - & - \\
        \num{51} & \num{4.902} & [1, 9, 5, 2, 4] & \num{0.968} & \num{134.698} & \num{0.706} & \num{1.187} & \num{139.517} & \num{0.562}\\
        \num{52} & \num{4.808} & [1, 4, 5, 9, 4] & \num{1.029} & \num{136.641} & \num{0.726} & \num{0.939} & \num{139.813} & \num{0.351}\\
        \num{53} & \num{4.717} & [1, 2, 1, 1, 7] & \num{0.938} & \num{135.830} & \num{0.128} & \num{0.943} & \num{139.626} & \num{0.484}\\
        \num{54} & \num{4.629} & [1, 1, 1, 2, 3] & \num{1.407} & \num{135.436} & \num{0.162} & \num{1.057} & \num{141.601} & \num{0.923}\\
        \num{55} & \num{4.545} & [1, 1, 5, 55, 2] & \num{1.083} & \num{137.088} & \num{1.056} & \num{1.735} & \num{141.899} & \num{1.136}\\
        \num{56} & \num{4.464} & [2, 6, 2, 8, 3] & \num{1.178} & \num{135.970} & \num{0.232} & \num{1.194} & \num{139.089} & \num{0.867}\\
        \num{57} & \num{4.386} & [2, 1, 1, 2, 4] & \num{1.517} & \num{135.657} & \num{0.001} & \num{1.093} & \num{139.586} & \num{0.513}\\
        \num{58} & \num{4.310} & [3, 4, 2, 7, 1] & \num{1.116} & \num{136.368} & \num{0.525} & \num{1.587} & \num{142.270} & \num{1.400}\\
        \num{59} & \num{4.237} & [4, 4, 1, 1, 1] & \num{1.160} & \num{135.943} & \num{0.212} & \num{1.242} & \num{140.377} & \num{0.051}\\
        \num{60} & \num{4.167} & [6, 203, 1, 1, 3] & \num{1.345} & \num{127.712} & \num{5.856} & \num{1.784} & \num{136.275} & \num{2.873}\\
        \bottomrule
    \end{tabular}
    \label{tab:sparse_mesh_50}
\end{table}

\begin{table}
    \centering
    \caption{Results for the sparse mesh solution using approximately 250 subintervals.}
    \begin{tabular}{ccccccccc}
        \toprule
        $N$ & $\rho$ & continued fraction of $\rho$ & $T_c$ (\unit{\second}) & $J_c$ (\unit{\kilo\gram}) & $E_c$ (\unit{\percent}) & $T_{J_2}$ (\unit{\second}) & $J_{J_2}$ (\unit{\kilo\gram}) & $E_{J_2}$ (\unit{\percent}) \\
        \midrule
        \num{245} & \num{1.020} & [49, 12, 2, 4, 1, 5] & \num{2.246} & \num{135.584} & \num{0.053} & \num{2.597} & \num{140.697} & \num{0.279}\\
        \num{246} & \num{1.016} & [61, 1, 1, 1, 2, 14] & \num{2.033} & \num{135.793} & \num{0.101} & \num{2.202} & \num{140.901} & \num{0.424}\\
        \num{247} & \num{1.012} & [82, 1, 1, 3, 1, 3] & \num{1.511} & \num{136.694} & \num{0.766} & \num{2.245} & \num{141.460} & \num{0.823}\\
        \num{248} & \num{1.008} & [124, 1, 1, 26, 2, 4] & \num{2.847} & \num{139.874} & \num{3.109} & \num{2.615} & \num{144.050} & \num{2.669}\\
        \num{249} & \num{1.004} & [251, 18, 1, 1, 1, 1] & \num{5.101} & \num{242.583} & \num{78.822} & Failed & - & - \\
        \num{250} & \num{1.000} & [1, 30559, 4, 5, 1, 8] & Failed & - & - & Failed & - & - \\
        \num{251} & \num{0.996} & [1, 247, 1, 26, 3, 1] & \num{2.970} & \num{210.737} & \num{55.347} & \num{2.296} & \num{198.140} & \num{41.221}\\
        \num{252} & \num{0.992} & [1, 124, 2, 18, 3, 44] & \num{1.714} & \num{139.469} & \num{2.810} & \num{2.565} & \num{143.372} & \num{2.186}\\
        \num{253} & \num{0.988} & [1, 83, 9, 1, 1, 1] & \num{2.761} & \num{136.450} & \num{0.585} & \num{2.686} & \num{140.830} & \num{0.374}\\
        \num{254} & \num{0.984} & [1, 62, 2, 1, 2, 3] & \num{2.313} & \num{135.597} & \num{0.043} & \num{5.137} & \num{139.947} & \num{0.255}\\
        \num{255} & \num{0.980} & [1, 49, 1, 11, 249, 35] & \num{2.593} & \num{135.402} & \num{0.187} & \num{3.050} & \num{139.801} & \num{0.360}\\
        \bottomrule
    \end{tabular}
    \label{tab:sparse_mesh_250}
\end{table}

\begin{table}
    \centering
    \caption{Results for the sparse mesh solution using approximately 410 subintervals.}
    \begin{tabular}{ccccccccc}
        \toprule
        $N$ & $\rho$ & continued fraction of $\rho$ & $T_c$ (\unit{\second}) & $J_c$ (\unit{\kilo\gram}) & $E_c$ (\unit{\percent}) & $T_{J_2}$ (\unit{\second}) & $J_{J_2}$ (\unit{\kilo\gram}) & $E_{J_2}$ (\unit{\percent}) \\
        \midrule
        \num{400} & \num{0.625} & [1, 1, 1, 1, 1, 763] & \num{3.269} & \num{136.016} & \num{0.265} & \num{3.438} & \num{140.877} & \num{0.407}\\
        \num{401} & \num{0.623} & [1, 1, 1, 1, 1, 9] & \num{3.030} & \num{135.708} & \num{0.039} & \num{3.760} & \num{140.472} & \num{0.119}\\
        \num{402} & \num{0.622} & [1, 1, 1, 1, 1, 4] & \num{2.945} & \num{135.736} & \num{0.059} & \num{9.670} & \num{140.442} & \num{0.097}\\
        \num{403} & \num{0.620} & [1, 1, 1, 1, 1, 2] & \num{3.359} & \num{135.664} & \num{0.006} & \num{4.256} & \num{140.407} & \num{0.072}\\
        \num{404} & \num{0.619} & [1, 1, 1, 1, 1, 1] & \num{4.929} & \num{135.684} & \num{0.021} & \num{3.569} & \num{140.424} & \num{0.085}\\
        \num{405} & \num{0.617} & [1, 1, 1, 1, 1, 1] & \num{2.739} & \num{135.676} & \num{0.015} & \num{4.971} & \num{140.428} & \num{0.087}\\
        \num{406} & \num{0.616} & [1, 1, 1, 1, 1, 1] & \num{3.357} & \num{135.781} & \num{0.093} & \num{3.271} & \num{140.440} & \num{0.096}\\
        \num{407} & \num{0.614} & [1, 1, 1, 1, 2, 4] & \num{3.146} & \num{135.611} & \num{0.033} & \num{4.119} & \num{140.365} & \num{0.043}\\
        \num{408} & \num{0.613} & [1, 1, 1, 1, 2, 1] & \num{2.599} & \num{135.662} & \num{0.005} & \num{2.852} & \num{140.369} & \num{0.045}\\
        \num{409} & \num{0.611} & [1, 1, 1, 1, 2, 1] & \num{3.412} & \num{135.608} & \num{0.035} & \num{3.400} & \num{140.417} & \num{0.079}\\
        \num{410} & \num{0.610} & [1, 1, 1, 1, 3, 1] & \num{3.163} & \num{135.697} & \num{0.031} & \num{3.072} & \num{140.373} & \num{0.048}\\
        \num{411} & \num{0.608} & [1, 1, 1, 1, 4, 4] & \num{4.056} & \num{135.644} & \num{0.009} & \num{3.209} & \num{140.356} & \num{0.036}\\
        \num{412} & \num{0.607} & [1, 1, 1, 1, 5, 3] & \num{3.476} & \num{135.693} & \num{0.027} & \num{4.194} & \num{140.369} & \num{0.045}\\
        \num{413} & \num{0.605} & [1, 1, 1, 1, 6, 1] & \num{2.896} & \num{135.633} & \num{0.017} & \num{3.025} & \num{140.326} & \num{0.014}\\
        \num{414} & \num{0.604} & [1, 1, 1, 1, 9, 1] & \num{4.232} & \num{135.824} & \num{0.124} & \num{3.547} & \num{140.351} & \num{0.032}\\
        \num{415} & \num{0.602} & [1, 1, 1, 1, 16, 7] & \num{3.200} & \num{135.714} & \num{0.043} & \num{3.437} & \num{140.191} & \num{0.082}\\
        \num{416} & \num{0.601} & [1, 1, 1, 1, 41, 1] & \num{4.114} & \num{136.929} & \num{0.939} & \num{3.929} & \num{140.102} & \num{0.145}\\
        \num{417} & \num{0.600} & [1, 1, 2, 79, 1, 2] & \num{4.253} & \num{136.863} & \num{0.890} & \num{13.056} & \num{142.329} & \num{1.442}\\
        \num{418} & \num{0.598} & [1, 1, 2, 20, 3, 2] & \num{3.272} & \num{135.768} & \num{0.083} & \num{3.722} & \num{140.900} & \num{0.424}\\
        \num{419} & \num{0.597} & [1, 1, 2, 11, 1, 1] & \num{3.527} & \num{135.723} & \num{0.050} & \num{2.694} & \num{140.616} & \num{0.221}\\
        \num{420} & \num{0.595} & [1, 1, 2, 7, 1, 28] & \num{2.872} & \num{135.663} & \num{0.005} & \num{3.071} & \num{140.578} & \num{0.194}\\
        \bottomrule
    \end{tabular}
    \label{tab:sparse_mesh_410}
\end{table}

Analyzing the results for approximately 50 subintervals, it can be firstly observed that the worst performance occurs with exactly 50 subintervals. With $\rho$ near an integer, the optimization routine for the discretized problem fails to converge in both cases with or without the $J_2$ perturbation. This occurs because all mesh points are located at the same position on the unit circle, losing most of the dynamic information. Two other cases, 40 and 60 subintervals, are also problematic. Despite the convergence of the discretized problem, they yield higher relative errors because $\rho$ is close to $\tfrac{25}{4}$ and $\tfrac{25}{6}$, which are rational numbers with small denominators 4 and 6. Meanwhile, for $N = 43, 54, 57,$ and $59$, whose continued fraction expansions have all terms below 5, the results show relative errors under \qty{1}{\percent} within about 1 second of computational time. These findings align with Section \ref{sec:irrational}, indicating that a strongly irrational $\rho$ distributes mesh points uniformly on the unit circle and leads to better convergence and accuracy. They also demonstrate that the proposed method solves complex multi-revolution low-thrust transfers in just a few seconds, which is a promising result for academic and practical applications. 

\begin{figure}
    \centering
    \includegraphics[width=0.8\textwidth]{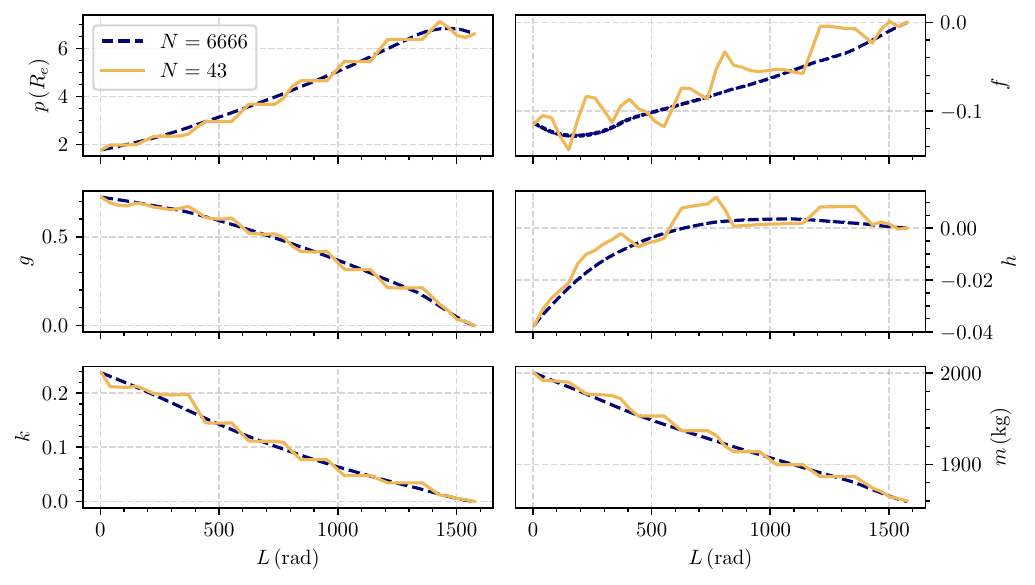}
    \caption{The comparison of the state variables for $N=43$ and $N = 6666$ with $J_2$ perturbation. The results given by the very sparse mesh oscillate around the reference solution.}
    \label{fig:compare_states}
\end{figure}

A comparison of the state variables for $N=43$ and $N=6666$ with $J_2$ perturbation is shown in Figure~\ref{fig:compare_states}. The results obtained with the very sparse mesh ($N=43$) oscillate around the reference solution ($N=6666$), indicating the correctness of the solution and the effectiveness of the proposed method. The effect of reordering the integration, as introduced in Section~\ref{sec:commutativity}, is also evident in the figure.
In low-thrust transfers, the solution typically exhibits about two coasting arcs and two burning arcs per revolution. During the coasting arcs, all six state variables remain nearly constant, appearing almost horizontal in the figure. In contrast, during the burning arcs, the state variables undergo secular variations and display nonzero slopes. As a result, the alternating coasting and burning arcs produce a stepped shape in the figure. For the $N=6666$ case, the steps are very small and invisible due to the large number of revolutions, whereas for the $N=43$ case, the steps are large and noticeable.
As noted in Section~\ref{sec:slow_dynamics}, for a given true longitude $L$, if the optimal control at that point is a coast, then the optimal control at points $L \pm 2k\pi$ (for a small integer $k$) will also be a coast in most cases, and vice versa for burning arcs. The near-commutativity of the vector field, as discussed in Section~\ref{sec:commutativity}, allows the consolidation of many coasting or burning points at $L \pm 2k\pi$ into a single integration step, which is fundamental to the effectiveness of the proposed method. This consolidation groups many coasting or burning phases into one step, explaining the multiple large steps observed in the $N=43$ case.

The experiments with approximately 250 subintervals illustrate poor performance when the rotation number $\rho$ is near 1. The optimization either fails to converge or yields large errors when the discretization has approximately one subinterval per revolution. In contrast, with approximately 410 subintervals, $\rho$ approaches the golden ratio $1/\phi \approx 0.618$, which is strongly irrational. All cases in this range yield accurate solutions in a few seconds, except for $N=400$ and $N=417$, where $\rho$ is close to the rational numbers $\tfrac{5}{8}$ and $\tfrac{3}{5}$. These results again demonstrate the effectiveness of the proposed method and underscore the importance of choosing subinterval lengths that are strongly irrational.

\begin{figure}
    \centering
    \includegraphics[width=0.8\textwidth]{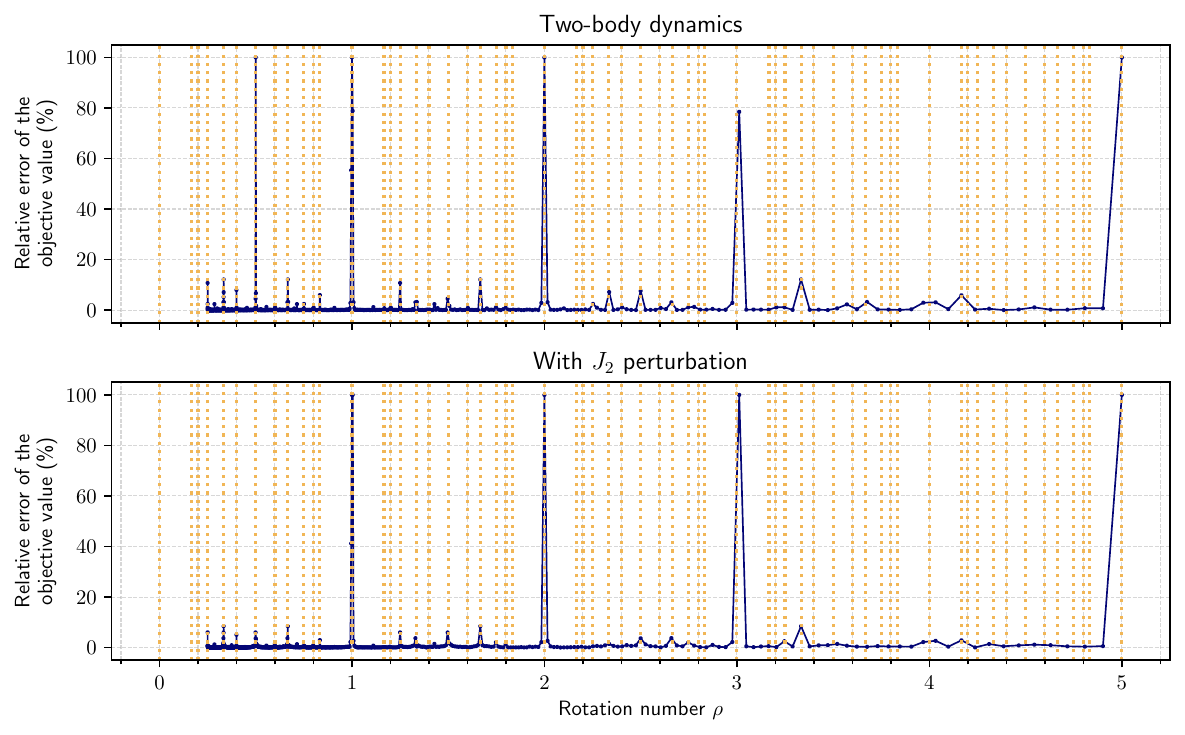}
    \caption{Relative error versus rotation number $\rho$ for $50 \leq N \leq 1000$. Orange lines indicate rational numbers with denominators $\leq 6$.}
    \label{fig:resonance}
\end{figure}

Figure~\ref{fig:resonance} presents a comprehensive analysis of the relationship between the rotation number $\rho$ and solution accuracy. The plot shows the relative errors of the objective function for all values of $N$ in the range $50 \leq N \leq 1000$, represented by the blue line. When the optimization fails to converge, the relative error is set to 100\%. The orange vertical lines indicate rational numbers with denominators no greater than 6. The results clearly demonstrate that relative errors increase significantly when $\rho$ approaches rational numbers with small denominators. Conversely, when $\rho$ is sufficiently distant from these rational values, the relative errors remain small, confirming the theoretical analysis presented in Section~\ref{sec:rationale}.

\subsubsection{Randomized Mesh Points}
\begin{figure}
    \centering
    \includegraphics[width=0.8\textwidth]{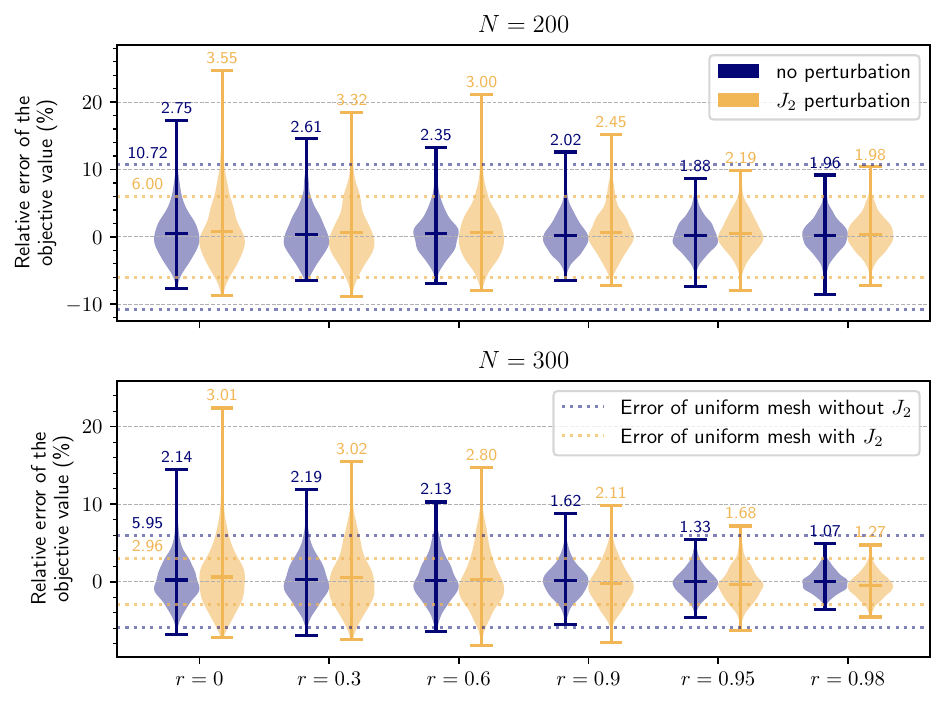}
    \caption{Results of the sparse mesh solution using randomized mesh points. The distribution of the relative deviations of the objective is shown in violin plots. }
    \label{fig:random_res}
\end{figure}

The randomized mesh points method is tested using 200 and 300 subintervals, corresponding to approximately $1.25$ and $0.83$ subintervals per revolution, respectively. These two numbers are chosen so that they are not strongly irrational, meaning the first method's condition is not satisfied and the solution with uniform meshes will produce large errors. For each case, different autocorrelation coefficients $r = 0, 0.3, 0.6, 0.9, 0.95,$ and $0.98$ are employed, and each case is run 1000 times with random seeds set to $0, 1, \ldots, 999$. Figure~\ref{fig:random_res} shows the distributions of the relative deviations of the objective with respect to the reference result in violin plots. The bars indicate the maximum, minimum, and mean values of the relative deviations. The numbers above the bars denote the mean of the absolute values of the relative deviations (the relative errors). Long dotted horizontal lines indicate the relative errors of the unperturbed uniform meshes (and their negative values), with values shown on the left. The unit of the relative error is \%.
The mean computational time is 2.72 seconds for 200 subintervals without $J_2$ perturbation, and 2.88 seconds with $J_2$ perturbation. For 300 subintervals, the mean computational time is 3.50 seconds without $J_2$ perturbation, and 3.78 seconds with $J_2$ perturbation. In all cases, the optimization process converges successfully.

As shown in the figure, the mean relative errors of the randomized cases with a high correlation parameter $r$ are significantly lower than those of the unperturbed uniform meshes. From the violin plots, it can also be observed that most results from the randomized cases lie within the range defined by the relative error (and its negative) of the unperturbed uniform meshes. This indicates that the randomization method is generally effective in improving solution accuracy. However, there are still some outliers with large relative errors, suggesting that the randomization method is not as robust as the strongly irrational subinterval length method. To use the randomization method in practice, it is recommended to run the optimization process multiple times with different random seeds and take the average of the results. This approach helps mitigate the impact of outliers and improve the overall accuracy of the solution.

Furthermore, the results show that introducing autocorrelation in the randomization process is vital. Cases with $r = 0$, which are equivalent to an i.i.d. random sequence, exhibit the worst performance in all scenarios. The performance continues to improve as $r$ increases, until $r = 0.95$ for 200 subintervals and $r = 0.98$ for 300 subintervals. Higher $r$ values lead to more evenly spaced mesh points, which benefits the solution's accuracy. However, excessively high $r$ values may not provide enough randomness, as evidenced in the $N=200, r=0.98$ case without the $J_2$ perturbation, where the mean relative error is larger than that for $r=0.95$. Generally, the more subintervals there are, the less randomness is needed to cover the unit circle, which explains why the optimal autocorrelation coefficient is higher for 300 subintervals than for 200.

Overall, the randomized mesh points method improves solution accuracy on average for multi-revolution low-thrust transfers when the strongly irrational subinterval length condition is not satisfied. However, it is generally less robust and less accurate than the strongly irrational rotation method when that condition is met. For the randomized method, the choice of autocorrelation coefficient is crucial and should be adjusted according to the specific problem and the number of subintervals.

\section{Conclusion}\label{sec:conclusion}
Multi-revolution low-thrust trajectory optimization problems are both essential and challenging in astrodynamics. This paper presents an efficient and accurate method for numerically solving these problems. The method's broad applicability to various objectives and practical perturbations is demonstrated through theoretical analysis and numerical experiments with challenging setups. Based on the Sundman transformation with pseudospectral methods, the proposed approach focuses on selecting a sparse mesh that ensures monotonicity, near-uniform spacing, and even sampling of the unit circle $\mathbb{R}/2\pi\mathbb{Z}$. Two methods are introduced to construct such sparse meshes: the strongly irrational rotation method and the randomized mesh points method. The first method is deterministic, relying on an irrational rotation number that is sufficiently distant from rational numbers with small denominators. A practical way to assess irrationality is provided by examining the continued fraction expansion of a real number. The second method is stochastic, achieved by perturbing mesh points with an autocorrelated random sequence. The underlying mechanism of the methods is discussed, revealing key insights into the dual roles of the mesh points as both integration points and sampling points. The effectiveness of the proposed methods is also attributed to the system's slow dynamics at equivalent true longitudes and the near-commutative nature of the vector fields.

To evaluate the performance of the proposed method, an elliptical, non-planar, multi-revolution low-thrust minimum-fuel rendezvous problem is examined under both two-body dynamics and with the inclusion of the $J_2$ perturbation. Results show that the proposed methods solve the test problem in a few seconds with high accuracy, marking a significant improvement over existing approaches. The strongly irrational subinterval length method is more robust and accurate, achieving relative errors below 0.1\% with carefully chosen subintervals. The randomized mesh points method is less robust and accurate due to inherent randomness, yet it maintains a mean relative error below 2\% and can be improved further by averaging multiple random runs. Numerical experiments also emphasize the critical importance of subinterval length selection and the autocorrelation coefficient, two core parameters that strongly influence the effectiveness of these methods. The results demonstrate that to achieve satisfactory accuracy, the subinterval length $h$ should be chosen such that $\rho = \frac{h}{2\pi}$ is strongly irrational, and the autocorrelation coefficient $r$ should be set to a value that balances randomness and uniformity, with optimal values depending on the number of subintervals and the specific problem.

\section*{Acknowledgments}
This work was supported by the National Natural Science Foundation of China (Grant No.12472355).

\bibliography{Resonance.bib, ref_nodes.bib}

\end{document}